\documentclass[onefignum,onetabnum]{siamart171218}

\usepackage{lipsum}
\usepackage{amsfonts}
\usepackage{graphicx}
\usepackage{epstopdf}
\usepackage{algorithmic}
\ifpdf
  \DeclareGraphicsExtensions{.eps,.pdf,.png,.jpg}
\else
  \DeclareGraphicsExtensions{.eps}
\fi

\newsiamremark{remark}{Remark}
\newsiamremark{hypothesis}{Hypothesis}
\crefname{hypothesis}{Hypothesis}{Hypotheses}
\newsiamthm{claim}{Claim}

\headers{Drift Behavior in a Bounded-Confidence Model with Media}{Oliver Zheng, Mason A. Porter}

\title{Drift Behavior in a Bounded-Confidence Opinion Model with Media Influence}

\author{
Oliver Zheng\thanks{Department of Mathematics, University of California, Los Angeles (UCLA), Los Angeles, CA {90095}, USA. E-mail: oliverz315@ucla.edu} 
\and
Mason A. Porter\thanks{{Department of Mathematics, University of California, Los Angeles (UCLA), Los Angeles, CA {90095}, USA; Department
of Sociology, University of California, Los Angeles (UCLA), Los Angeles, CA 90095{,} USA; 
Nordita (Nordic Institute for Theoretical Physics), KTH Royal Institute of Technology, Stockholm University, and Uppsala University, Stockholm, 106 91, Sweden;
Department of Mathematics, Stockholm University, Stockholm, 106 91, Sweden; {Santa Fe} Institute, Santa Fe, NM 87501{,} USA;
Digital Futures, Stockholm, 100 44, Sweden. E-mail: mason.porter@su.se}}
}

\usepackage{amsopn}

\makeatletter
\newcommand*{\addFileDependency}[1]{
  \typeout{(#1)}
  \@addtofilelist{#1}
  \IfFileExists{#1}{}{\typeout{No file #1.}}
}
\makeatother

\newcommand{\sech}{\operatorname{sech}}

\usepackage{enumitem}
\usepackage{amssymb}
\usepackage{graphicx}
\usepackage{subcaption}
\usepackage{siunitx}
\begin{document}

\maketitle

\begin{abstract}
People's opinions can change both from their interactions with each other and from their interactions with media sources. Bounded-confidence models (BCMs) of opinion dynamics provide one framework to study such dynamics. In a BCM, the nodes of a network are agents with continuous-valued opinions, and these agents interact with each other via the edges of the network. In this paper, we extend the original Deffuant--Weisbuch (DW) BCM by incorporating influence from two media sources --- one with a positive value and one with a negative value --- to capture the effects of a polarized media landscape. We show both numerically and analytically that our extended DW model exhibits drifting behavior in which a large cluster of opinions shifts toward one of the media agents. We analyze how the drift trajectory and speed depend on the model parameters, and we identify conditions in which drift is promoted or suppressed. Our results provide insight into how competing media sources can influence collective opinion formation in social systems. 
\end{abstract}






\section{Introduction} \label{intro}

People adjust their opinions on matters ranging from everyday decisions to complex social and political issues. In studies of opinion dynamics, researchers mathematically model and analyze how these opinions evolve with time~\cite{caldarelli2025,noorazar2020classical,starnini2025}. However, it is complicated to understand opinion dynamics, as the evolution of opinions is a complex, high-dimensional process that is influenced by many factors~\cite{sirbu2016opinion_chapter}. 

To study opinion dynamics in a controlled setting, it is useful to examine idealized mathematical models that capture a few essential mechanisms of opinion evolution. One common approach is to use agent-based models (ABMs)~\cite{volkening2025}, in which each agent is represented by a node of a network and possesses a distinct opinion in some opinion space. The edges of the network encode which agents are able to interact with each other. Studying models of opinion dynamics can provide insight into how small, local interactions lead to collective behavior, such as the formation of ``clusters'' of nodes with similar opinions~\cite{noorazar2020classical,starnini2025}. Depending on a model's formulation and parameter values, a system may ultimately reach consensus (i.e., there is one dominant opinion cluster), polarization (i.e., there are two major opinion clusters), or fragmentation (i.e., there are three or more major opinion clusters)~\cite{lorenz2007continuous}.

Agent-based models are often designed as idealized thought experiments to examine real-world scenarios~\cite{railsback2019agent}. For example, empirical research in the social sciences has demonstrated that individuals are more likely to interact with other individuals whose opinions are similar to theirs~\cite{stroud2011niche}, and it is desirable to incorporate this principle of ``homophily''~\cite{mcpherson2001birds} into ABMs. This observation motivates so-called \emph{bounded-confidence models} (BCMs)~\cite{deffuant2000mixing,hegselmann2002opinion,noorazar2020classical}, in which agents interact with each other and then compromise their opinions if and only if the difference between their opinions is within a ``confidence bound''. This compromise mechanism is reminiscent of the psychological idea of selective exposure, which describes the tendency of people to seek information and interactions that support
their existing views and to avoid those that challenge their views~\cite{chandler2011}. Under this assumption, an agent's views are influenced
directly only by agents or media sources with sufficiently similar views. The two foundational BCMs are the Deffuant--Weisbuch (DW) model~\cite{deffuant2000mixing} and the Hegselmann--Krause (HK) model~\cite{hegselmann2002opinion}. Both the DW model and the HK model evolve in discrete time. In the DW model, agents update their opinions asynchronously. At each time, one randomly (typically uniformly at random) selects a pair of adjacent agents to interact, and these agents then compromise their opinions if the difference between their opinions is within the confidence bound. In the HK model, agents update their opinions synchronously. At each time, all of the agents in a network update their opinions based on the means of the opinions of their neighbors with sufficiently nearby opinions.

Standard BCMs capture pairwise (i.e., ``dyadic'') interactions between agents, but they do not account fully for the broader informational environment in which people's opinions evolve, especially in modern societies~\cite{caldarelli2025}. Nowadays, individuals are exposed to a wide range of media sources that can significantly influence their opinions~\cite{noorazar2020classical}. These media sources include both traditional outlets such as television, radio, and newspapers and digital platforms such as social media, blogs, and online news aggregators~\cite{boulianne2009internet,mcpherson2001birds}. Recent changes in the media landscape, including the elimination of the so-called ``fairness doctrine'' and the rise of social media, have yielded an ecosystem with increased polarization~\cite{bc2021,gentzkow2010media,prior2007post}. Such a polarized media landscape can shape public opinion by reinforcing or radicalizing existing views~\cite{barbera2015tweeting,sunstein2001republic}. However, although media exposure can influence individual opinions, the extent to which polarized media drives large-scale polarization is unclear, and some empirical evidence suggests that its aggregate effects may be smaller than is commonly assumed~\cite{hosseinmardi2021youtube}. Nevertheless, it is essential to understand the effects of polarized media to develop a complete explanation of the macroscopic landscape of opinion dynamics in human populations.

In an ABM, one can represent media sources as ``stubborn'' agents that affect other agents' opinions but never update their own opinions. One prominent ABM that includes stubborn nodes is the Friedkin--Johnsen (FJ) model~\cite{friedkin1990social}, which is an extension of the French--DeGroot (FD) consensus model~\cite{degroot1974reaching}. In the FJ model, each agent has a distinct compromise parameter that controls how much they are influenced by other agents. A large compromise-parameter value indicates that an agent is open to adjusting its opinion, whereas a small value indicates that an agent is stubborn and tends to retain its current opinion.

Since the development of the FJ model, many researchers have examined ABMs with stubborn agents, which they have interpreted in a variety of ways. Mobilia~\cite{mobilia2003zealot} extended the classical voter model~\cite{clifford1973model} by including a single stubborn agent that they called a ``zealot'', and they showed analytically that the zealot has a significant effect on the macroscopic behavior of a population. Xie et al.~\cite{xie2011social} extended the classical voter model by including a ``committed minority'' of stubborn agents that share the same opinion. Using numerical computations, they showed that sufficiently large committe minorities (with at least about 10\% of a population) cause the majority of the population to adopt the minority opinion. Brooks and Porter~\cite{brooks2020media} studied a BCM with stubborn media agents that spread information of different opinions and different qualities. Using both theoretical analysis and numerical simulations, Brooks et al.~\cite{brooks2024sigmoidal} investigated the emergence of polarization from competing media sources in a BCM with sigmoidal interactions between agents. Finally, and of particular interest to us, Pineda and Buend\'{i}a~\cite{pineda2015mass} (PB) extended both the DW and HK models by including a single stubborn media agent. The PB model, which we discuss further in Section~\ref{sec:background}, is an important inspiration for our {ABM}.

Our paper proceeds as follows. In Section~\ref{sec:background}, we discuss the DW and PB models and clarify how our {ABM} differs from these models. In Section~\ref{sec:model}, we give a precise mathematical formulation of our {ABM}, including its governing equations and parameters. In Section~\ref{sec:numerical_exploration}, we present results from numerical simulations to illustrate the qualitative behavior of our {ABM} in different regions of parameter space. Notably, these simulations reveal a drift behavior in which the system's opinion distribution drifts gradually in one direction over time rather than remaining centered or fluctuating symmetrically. In Section~\ref{sec:mfa}, we derive a {continuous-time} approximation of our {ABM} to develop a better understanding of its macroscopic behavior. In Section~\ref{sec:drift}, we use our {continuous-time} approximation to analyze opinion drift and compare our analytical predictions with our simulation results. Finally, in Section~\ref{sec:conclusions}, we summarize our main findings and discuss extensions of our work. Our simulation code and all scripts to generate our figures are available at \url{https://github.com/oliverz3233/BCM_with_Media_code}.


\section{Background}\label{sec:background}

In this section, we give mathematical descriptions of the DW and PB models.


\subsection{The Deffuant--Weisbuch (DW) Model}\label{subsec:dw}

The baseline DW model~\cite{deffuant2000mixing} is a BCM with asynchronous opinion updates in discrete time steps. Consider an unweighted and undirected network (i.e., graph) $G = (V, E)$. The nodes of $G$ represent agents with opinions in the interval $[-1, 1]$, and the edges of $G$ encode social and/or communication ties between agents. We denote the opinion of node $i$ at time $t$ by $x_i(t)$.

At each time $t$, we select two distinct nodes $i$ and $j$ uniformly at random. If their opinions are sufficiently close, which we quantify with the inequality 
\begin{equation}
	|x_i(t) - x_j(t)| < \epsilon \, ,
\end{equation}	
where $\epsilon \in (0, 2]$ is a confidence bound, then nodes $i$ and $j$ compromise their opinions according to the update rule
\begin{equation}
\label{eq:dw_update_rule}
	\begin{aligned}
		x_i(t + 1) &= x_i(t)  +  \mu \big[x_j(t) - x_i(t)\big] \, , \\
		x_j(t + 1) &= x_j(t)  +  \mu \big[x_i(t) - x_j(t)\big] \, ,
	\end{aligned}
\end{equation}
where $\mu \in (0, 0.5]$ is a compromise parameter, which determines how much agents adjust their opinions during a compromise. If the opinions of nodes $i$ and $j$ are not sufficiently close to each other, then they do not compromise, so $x_i$ and $x_j$ remain the same.


\subsection{The Pineda--Buend\'{i}a (PB) Model}
There are two versions of the PB model: one extends the DW model, and the {other extends} the HK model. In the present paper, we consider only the extension of the DW model. In this version of the PB model, opinions update asynchronously and there is a single stubborn agent with opinion $S$ that represents a media source. At time $t$, one selects an agent $i$ uniformly at random. With probability $p_M$, if $|x_i(t) - S| < \epsilon$, agent $i$ compromises with the media source according to the update rule
\begin{equation}
	x_i(t + 1) =  x_i(t) + \mu[S - x_i(t)] \, .
\end{equation}
Otherwise, with probability $1 - p_M$, one selects another agent $j$ uniformly at random, and then agents $i$ and $j$ follow the normal DW opinion update~\eqref{eq:dw_update_rule} if their opinions are sufficiently close to 
each other.\footnote{Selecting an agent $i$ uniformly at random and then selecting a second agent $j \neq i$ uniformly at random, as we do here, is not equivalent to selecting a uniformly random edge $(i,j)$ (and hence selecting a pair of distinct agents uniformly at random) from the set of all edges~\cite{li2023}. The former procedure samples from the set of ordered node pairs, whereas the latter procedure samples from the set of unordered node pairs. For undirected and unweighted complete graphs (which is the only type of network that we examine), both procedures select each unordered pair with the same probability.
 }


\section{Model Definition}\label{sec:model}

In our extended PB model, each node has an opinion in the interval $[-1, 1]$. Our {ABM} includes both a confidence bound $\epsilon \in (0, 2]$ and a compromise parameter $\mu$. We suppose that there are $N$ ordinary agents. To represent a polarized media landscape, {we introduce two additional} stubborn media agents with fixed opinions $M_-$ and $M_+$, where $M_- \in [-1, 0)$ lies in one half of the opinion space and $M_+ \in (0, 1]$ lies in the other half of opinion space. {Therefore, the total number of nodes in the system is $N + 2$.} These media agents never update their opinions. We refer to the media agent with opinion $M_-$ as the ``negative media agent'' and to the media agent with opinion $M_+$ as the ``positive media agent''.

To make our model easier to analyze, we make the following assumptions:
\begin{enumerate}
    \item[(1)]{The agents represent nodes in a complete graph $K_{N + 2}$, so any two agents can interact with each other.}
    \item[(2)]{We set the compromise parameter to $\mu = 0.5$ both for interactions between two normal agents and for interactions between a normal agent and a media agent.}
    \item[(3)]{We set $M_- = -M_+$, so the media agents' opinions exactly oppose each other. Henceforth, we denote the media opinions by $M$ and $-M$, where $M \in (0, 1]$.} 
\end{enumerate}

\begin{figure}[htbp]
    \centering
    \includegraphics[width=0.8\textwidth]{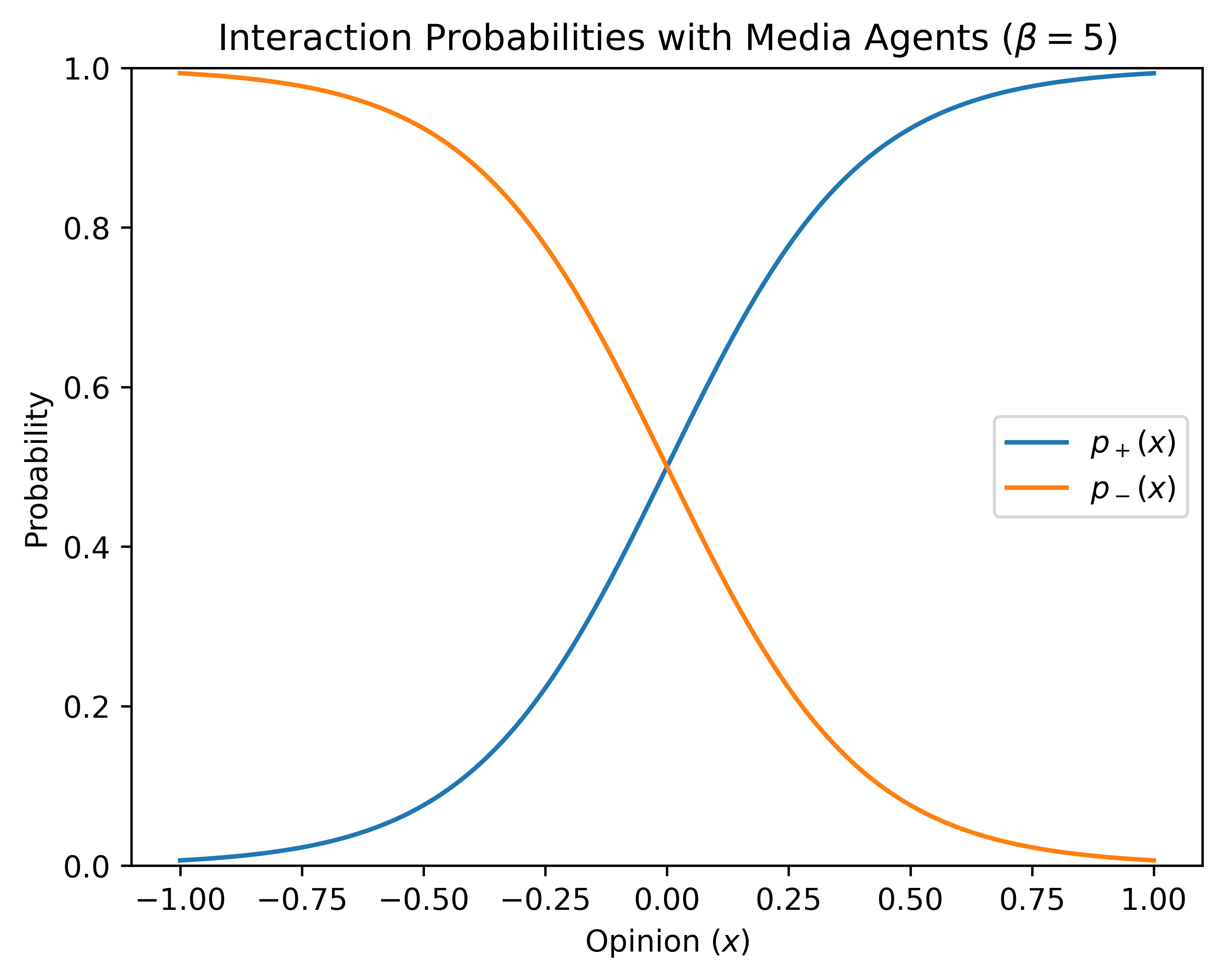}
    \caption{Media interaction probabilities as functions of opinion.}
    \label{fig:media_probs}
\end{figure}

At each time $t$, we select a node $i$ uniformly at random. With probability $p_M$ (which we set to $0.1$ unless we state otherwise), node $i$ interacts with one of the media agents. When this occurs, the probability that it interacts with the positive media agent is given by a logistic function $p_+(x)$ and the probability that it interacts with the negative media agent is $p_-(x) = 1 - p_+(x)$, where $x$ denotes the current opinion of node $i$. We let
\begin{equation}\label{eq:media_probs}
	\begin{aligned}
		p_+(x) &= \frac{1}{1 + e^{-\beta x}} \, , \\
		p_-(x) &= \frac{1}{1 + e^{\beta x}} \, .
	\end{aligned}
\end{equation} 
The parameter $\beta$ controls the steepness of the logistic function. Throughout this paper, we set $\beta = 5$. We show examples of these probability functions in Figure~\ref{fig:media_probs}.

\begin{figure}[htbp]
    \centering
    \includegraphics[width=\textwidth]{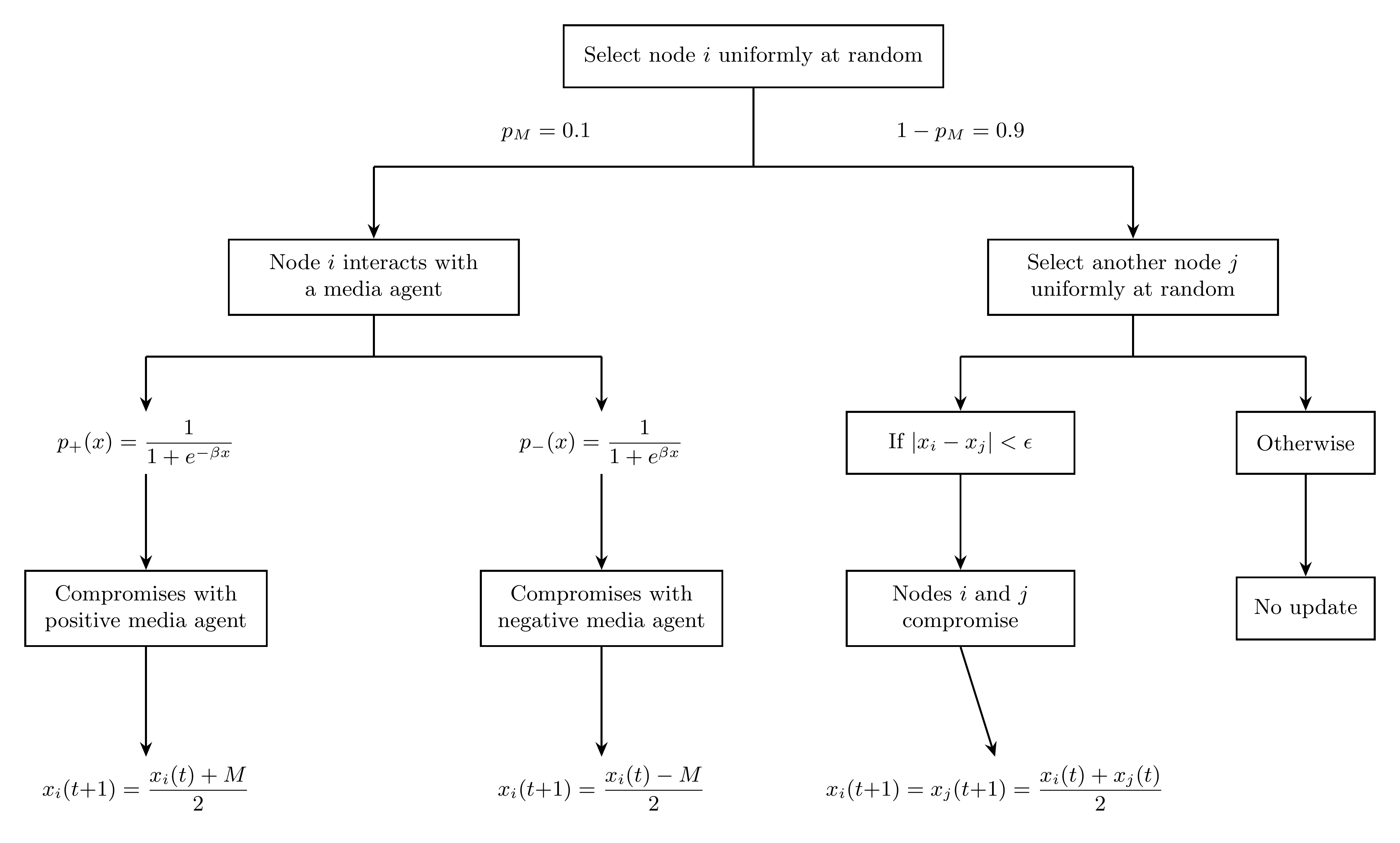}
    \caption{Schematic illustration of a (potential) opinion update in
    our {ABM} at each discrete time step $t$. We select a node $i$ uniformly at random. With probability $p_M$, node $i$ interacts with a media agent, where the probability of interacting with the positive media agent is 
    $p_+(x)$, and the probability of interacting with the negative media agent is 
    $p_-(x)$. With probability $1 - p_M$, node $i$ instead interacts with a uniformly randomly selected node $j$, and then nodes $i$ and $j$ update their opinions according to \eqref{eq:update_rule} if $|x_i - x_j| < \epsilon$.
    }
    \label{fig:model_diagram}
\end{figure}

If node $i$ does not interact with one of the media agents (i.e., with probability $1 - p_M = 0.9$), we select another node $j$ uniformly at random. {If} $|x_i(t) - x_j(t)| < \epsilon$, then nodes $i$ and $j$ update their opinions according to {the update rule
\begin{equation}
    \label{eq:update_rule}
    \begin{aligned}
        x_i(t + 1) &= \frac{x_i(t) + x_j(t)}{2} \, , \\
        x_j(t + 1) &= \frac{x_i(t) + x_j(t)}{2} \, .
    \end{aligned}
\end{equation}
Otherwise, if $|x_i(t) - x_j(t)| \geq \epsilon$,} their opinions stay the same. When $\mu = 0.5$ in the DW update rule~\eqref{eq:dw_update_rule}, it reduces to the update rule~\eqref{eq:update_rule}. In Figure~\ref{fig:model_diagram}, we show a schematic illustration of an opinion-update step in our {ABM}.

Our approach is inspired by the PB model, but it has two key changes:
\begin{itemize} 
\item[(1)]{Rather than a single stubborn agent, we include two stubborn agents, which we position at opposite opinion values to represent a polarized media landscape.} 
\item[(2)]{We allow agents to interact with the media agents regardless of the confidence bound. We do this to reflect the ubiquity of mass media and its ability to reach audiences with different opinions~\cite{nyhan2023facebook}. However, just as homophily influences the interactions between individuals, the agents also tend to prefer media sources that align with their existing views~\cite{stroud2011niche}. To account for this phenomenon, the probabilities in~\eqref{eq:media_probs} ensure that agents whose opinions are further from a particular media agent are less likely to interact with it.} 
\end{itemize}
These changes enable us to study how the presence of polarized and pervasive media can affect long-term opinion dynamics in a population.


\section{Numerical Exploration}\label{sec:numerical_exploration}

In this section, we present numerical results that illustrate the qualitative behavior of our ABM. In Section~\ref{subsec:without_media}, we examine the standard DW model (which does not include media agents) to establish a baseline. 
In Section~\ref{subsec:with_media}, we incorporate media agents and examine how their presence affects the long-term behavior of our ABM. In Section~\ref{subsec:numerical_drift}, we give numerical evidence of opinion drift 
and explore how the model parameters affect it. In all but one of our numerical simulations, we draw the initial agent opinions independently from a uniform distribution on the interval $[-1,1]$.


\subsection{Without Media Agents (i.e., $p_M = 0$)}\label{subsec:without_media}

\begin{figure}[htbp]
    \centering    
    \begin{subfigure}{0.48\textwidth}
        \centering
        \includegraphics[width=\linewidth]{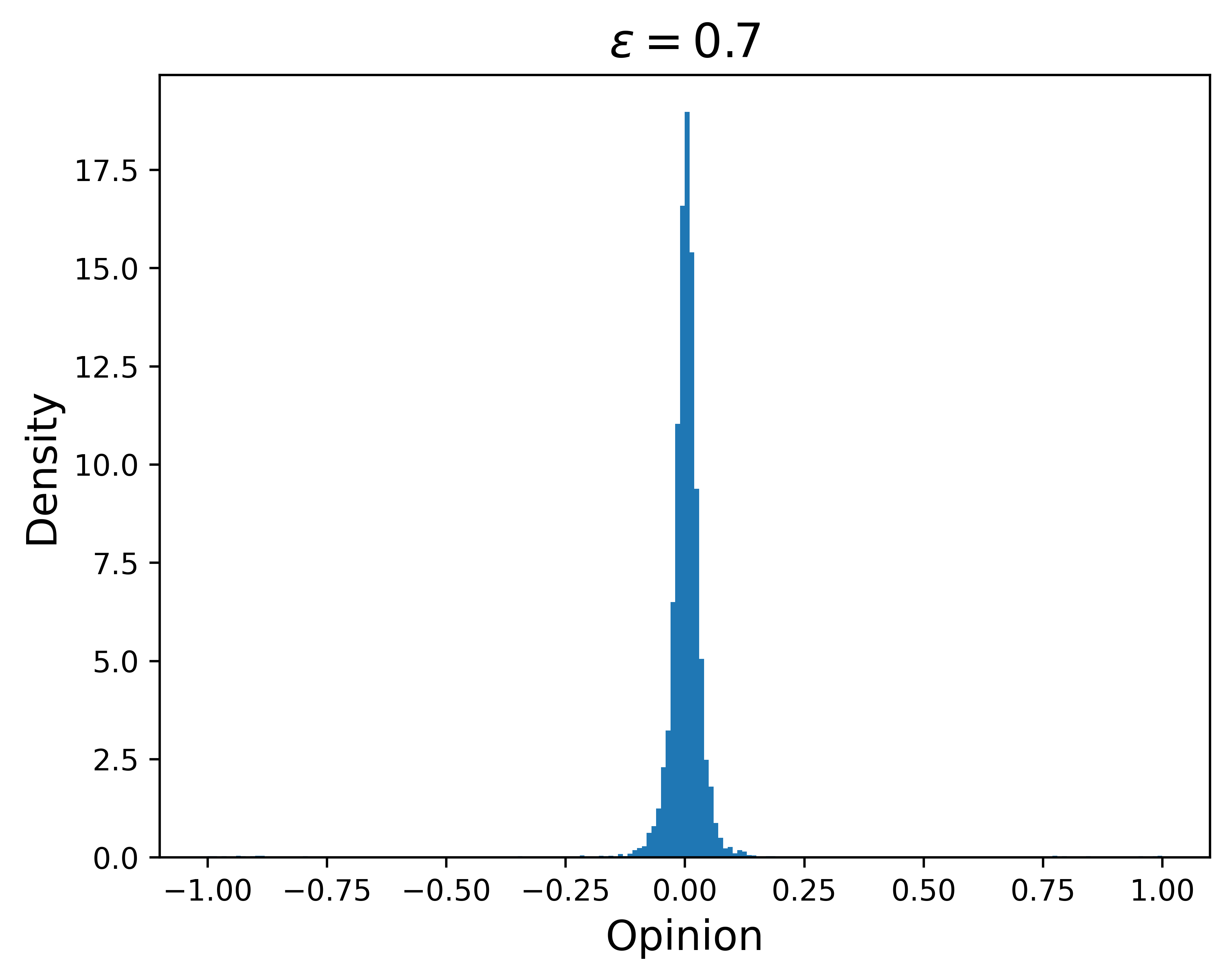}
        \caption{Consensus}
    \end{subfigure}
    \hfill
    \begin{subfigure}{0.48\textwidth}
        \centering
        \includegraphics[width=\linewidth]{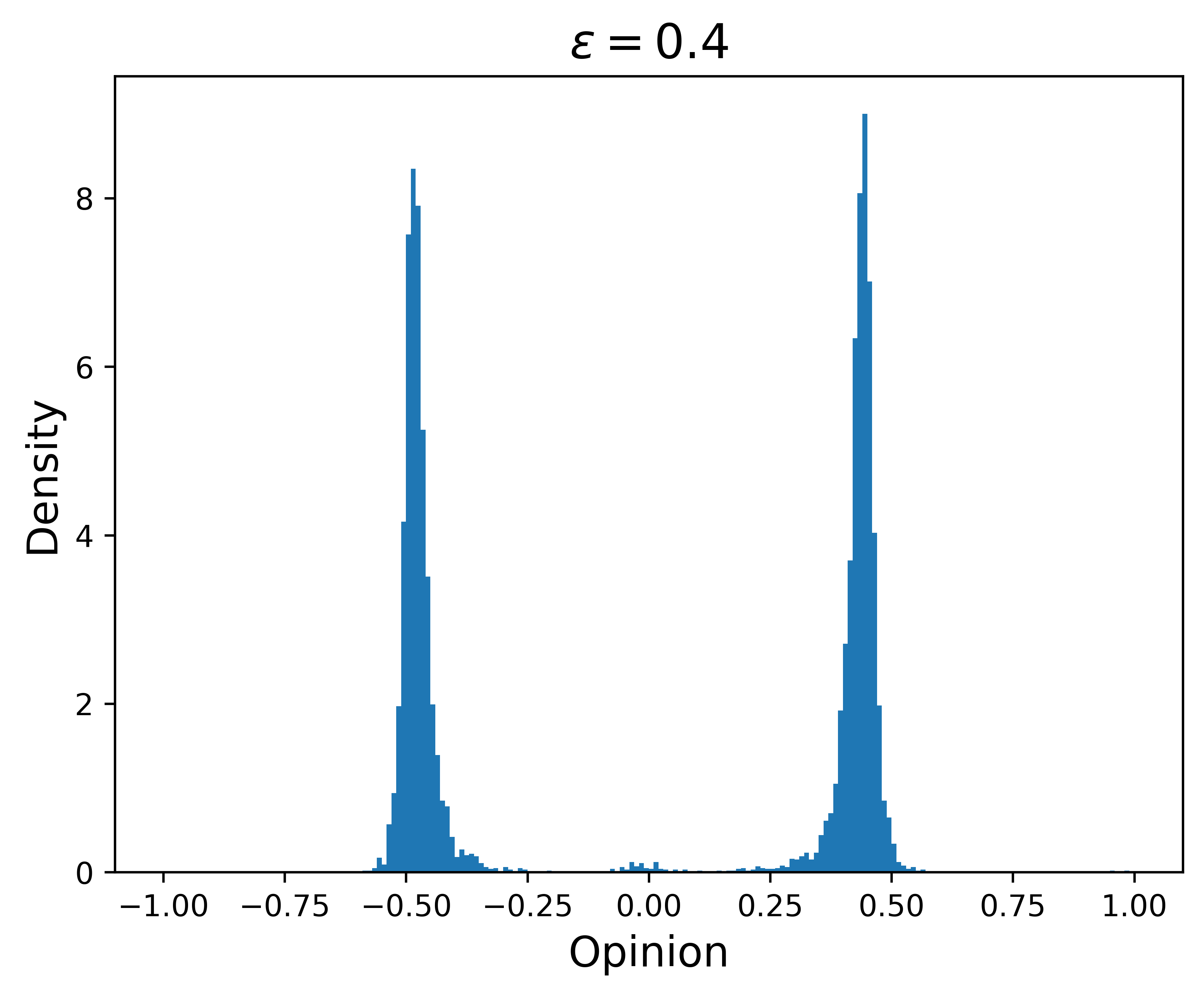}
        \caption{Polarization}
    \end{subfigure}
    \vspace{0.5em}
    \begin{subfigure}{0.48\textwidth}
        \centering
        \includegraphics[width=\linewidth]{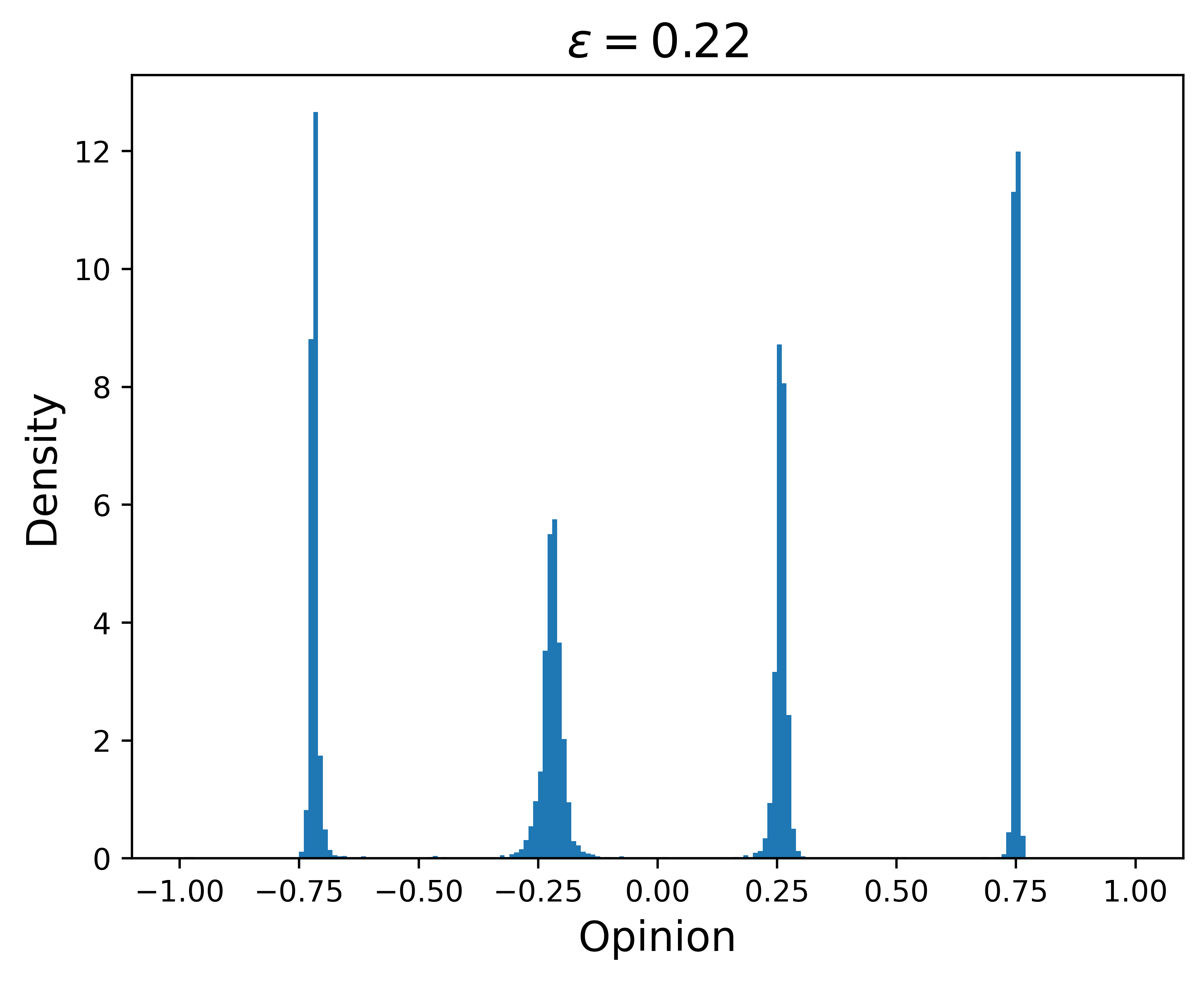}
        \caption{Fragmentation}
    \end{subfigure}
        \caption{Three different long-term outcomes of the baseline DW model. Each panel is a single simulation of the DW model on a complete graph with $N =  \num{10000}$ nodes. In these simulations, the confidence bound is (a) $\epsilon = 0.7$, (b) $\epsilon = 0.4$, and (c) $\epsilon = 0.22$. We stop each simulation after $\num{100000}$ time steps.}     
    \label{fig:long-term_outcomes}
\end{figure}

In our simulations of the standard DW model, we observe the following types of long-term outcomes:
\begin{enumerate}[label=(\arabic*)]
    \item{Consensus: There is one major opinion cluster. We refer to this cluster as the ``dominant'' opinion cluster.}
    \item{Polarization: There are exactly two major opinion clusters.}
    \item{Fragmentation: There are three or more major opinion clusters.}
\end{enumerate} 
In Figure~\ref{fig:long-term_outcomes}, we show examples of each of the three outcomes.

We describe an opinion cluster as a ``major'' cluster when its constituent agents, which have opinions that are close to each other, comprise a significant portion of a population. In practice, we identify opinion clusters using a simple one-dimensional clustering algorithm. We first sort all agent opinions in ascending order and then scan through them sequentially. We start with the agent with the smallest opinion, and we assign subsequent agents to the same opinion cluster when their opinion is within $0.1$ of the opinion of the first agent in that cluster. We then filter out clusters whose size is strictly less than $10$\% of the total population.

As is typical of studies of opinion models with continuous-valued opinions, our choices for determining opinion clusters and referring to certain opinion clusters as ``major" clusters include some arbitrariness. 
See, e.g.,~\cite{li2025adaptive,li2023} for other choices and associated discussions.

\begin{figure}[htbp]
    \centering
    \includegraphics[width=0.8\textwidth]{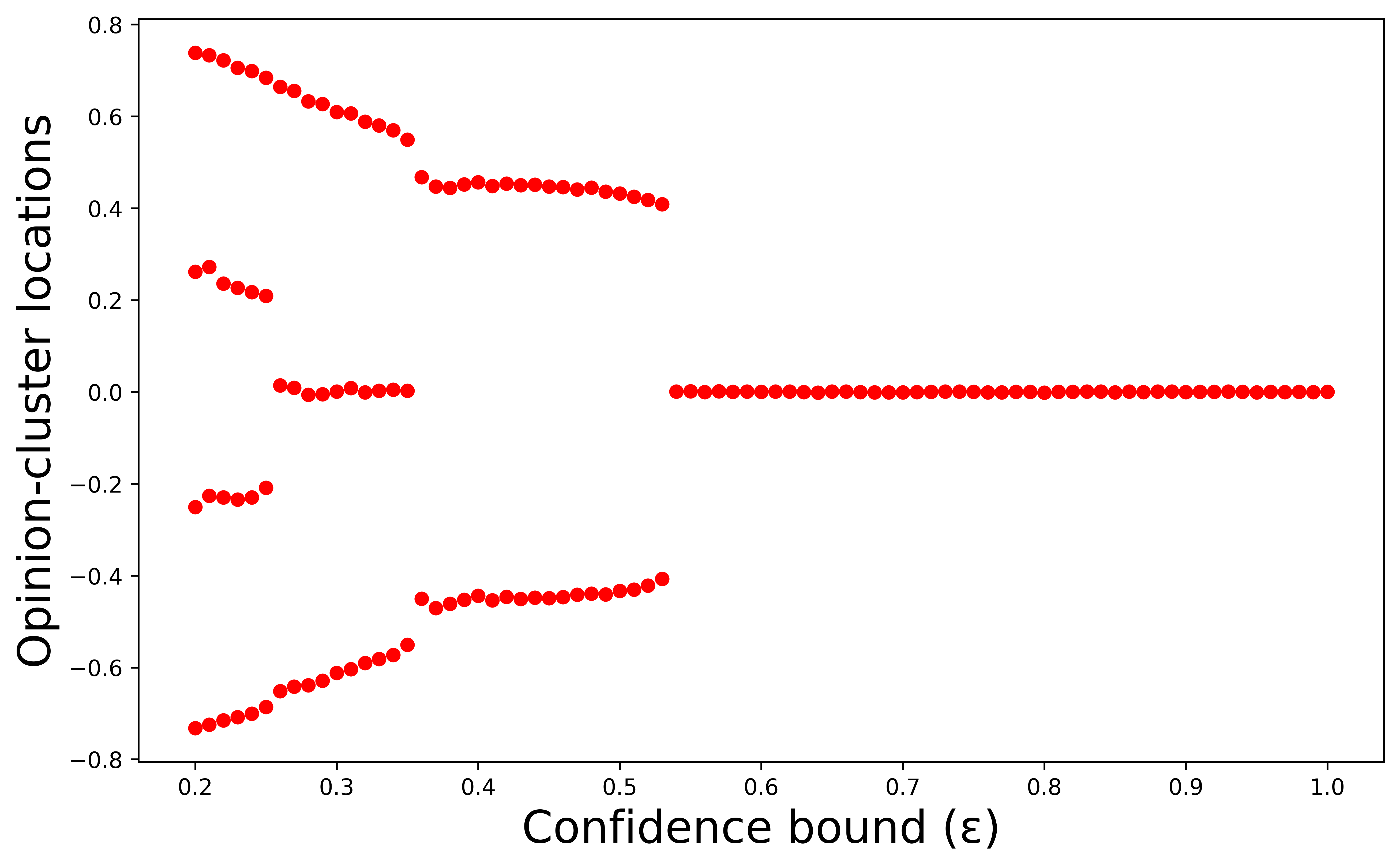}
    \caption{Bifurcation diagram of the baseline DW model. The horizontal axis is the confidence bound, and the vertical axis gives the locations of the major opinion clusters. We average each data point across 50 simulations, and we stop each simulation after $\num{1000000}$ time steps. For each simulation,
    we draw the initial opinions independently from a uniform distribution on the interval $[-1,1]$. 
    We omit confidence-bound values 
    $\epsilon < 0.2$ because the number of opinion clusters is large in this region of parameter space.
    }
    \label{fig:dw_bif_diagram}
\end{figure}

We visualize the relationship between the confidence bound $\epsilon$ and the number and locations of opinion clusters using a bifurcation diagram (see Figure~\ref{fig:dw_bif_diagram}), which shows the mean location of each opinion cluster as a function of $\epsilon$ across 50 simulations. For each simulation, we draw the initial opinions independently from a uniform distribution on the interval $[-1,1]$. We also choose the initial opinions in this way for all but one of our other numerical experiments.

Typically, as we increase $\epsilon$, we observe that the number of clusters decreases and that their locations move closer to the center $x = 0$. Deffuant et al.~\cite{deffuant2000mixing} observed that the number of opinion clusters is approximately
\begin{equation}\label{eq:num_clusters}
	n_c \approx \left\lfloor\frac{1}{\epsilon}\right\rfloor \, .
\end{equation}
Based on~\eqref{eq:num_clusters}, we expect fragmentation to occur when $0 < \epsilon \leq 1/3$, polarization to occur when $1/3 < \epsilon \leq 1/2$, and consensus to occur when $1/2 < \epsilon \leq 1$.

An important feature of the baseline DW model is the left--right symmetry of the mean long-term behavior. That is, if we let $x_k$ denote the opinion of the $k$th cluster, where $x_1 < x_2 <  \cdots < x_n$, we expect that
\begin{equation*}
	x_1 \approx -x_n \, , \; x_2 \approx -x_{n - 1} \, , \,  \ldots \,.
\end{equation*}
The asymmetries that we observe in Figure~\ref{fig:dw_bif_diagram} arise because the number of simulations is finite.

In Section~\ref{subsec:with_media}, we examine how the inclusion of media agents results in opinion drift, which breaks this symmetry.


\subsection{Inclusion of Media Agents (i.e., $p_M > 0$)}\label{subsec:with_media}

In this subsection, we set $p_M = 0.1$ and observe how the inclusion of media agents affects the long-term dynamics.

\begin{figure}[htbp]
    \centering
    \includegraphics[width=0.8\textwidth]{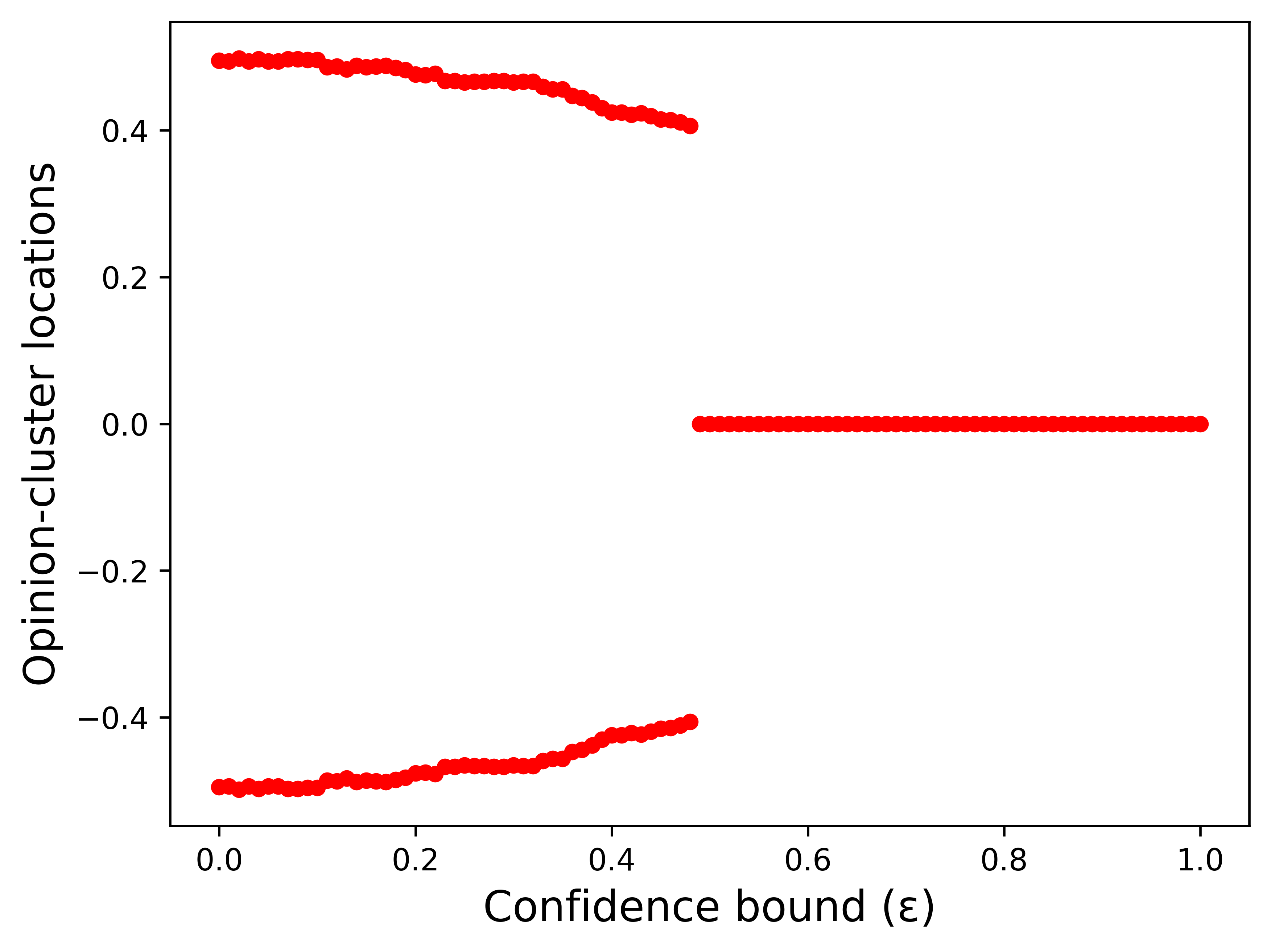}
    \caption{Bifurcation diagram of our ABM with media agents when the media opinion is $M = 0.5$. We average each data point across 50 simulations, and we stop each simulation after \num{1000000} time steps.
    }
    \label{fig:M=0.5_bif_diagram}
\end{figure}

With the media agents, we now have two adjustable parameters ($\epsilon$ and $M$). In Figure~\ref{fig:M=0.5_bif_diagram}, we show a bifurcation diagram for a fixed media opinion $M$.

In our simulations, the long-term behavior is either consensus or polarization; we never observe fragmentation.
 We again observe that the locations of the opinion clusters approach $0$ as we increase the confidence bound $\epsilon$. Intuitively, it makes sense that we never observe fragmentation.
  When $\epsilon$ is small, media effects dominate interactions between other agents and agents are pulled toward the two media agents, which hinders the formation of more than two stable opinion clusters.

\begin{figure}[htbp]
    \centering
    \includegraphics[width=0.8\textwidth]{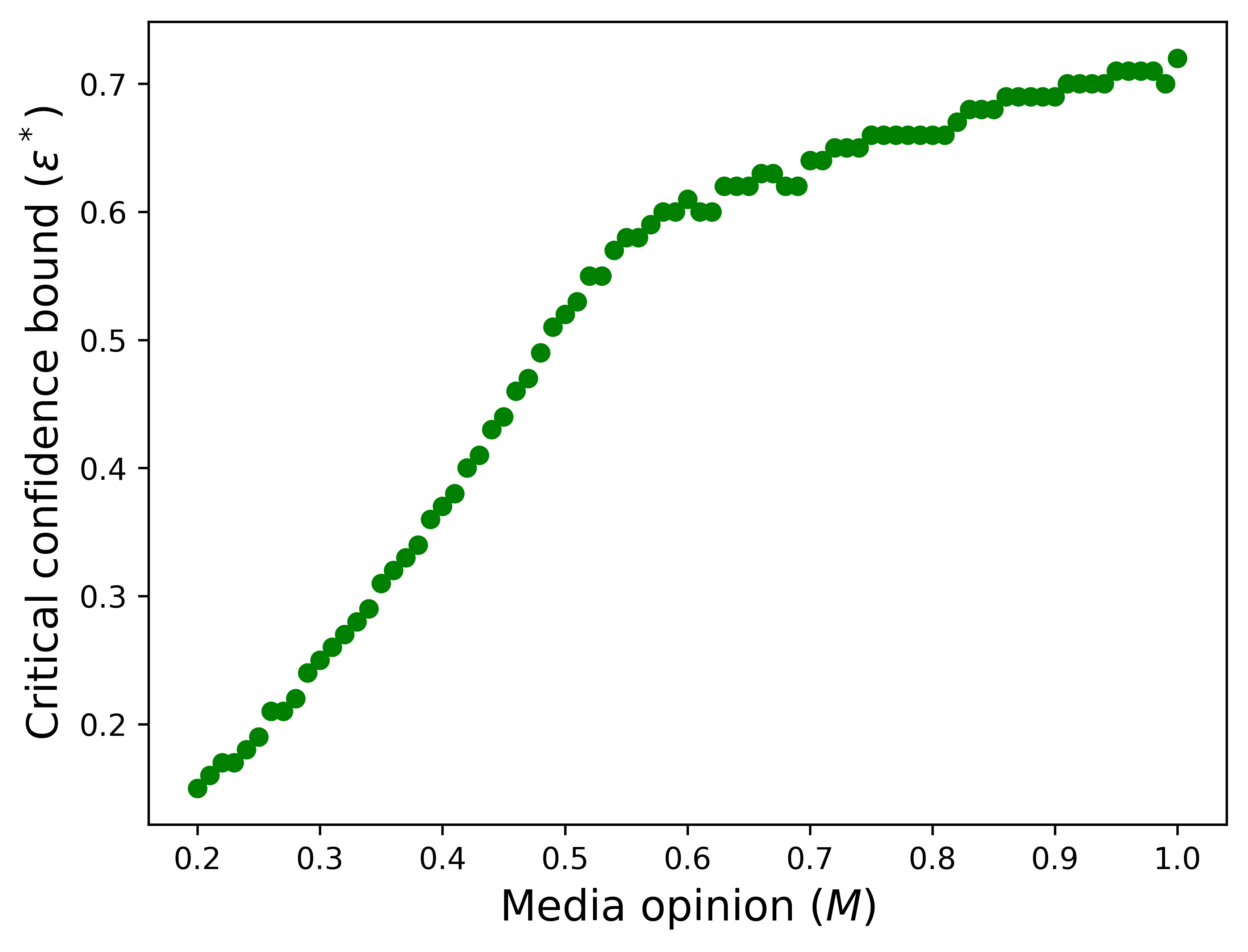}
    \caption{Critical confidence bound $\epsilon^*$ as a function of the media opinion $M$. Each data point represents one simulation. We omit media opinion values $M < 0.2$ because, in this parameter region, it is often unclear whether the final state is consensus or polarization.
    }
    \label{fig:eps_crit}
\end{figure}

For all values of the media opinion $M$, we also observe that there seems to be a ``critical confidence bound'' $\epsilon^*$. If $\epsilon < \epsilon^*$, we observe polarization; if $\epsilon > \epsilon^*$, we observe consensus. For instance, from Figure~\ref{fig:M=0.5_bif_diagram}, we deduce that $\epsilon^* \approx 0.5$. In Figure~\ref{fig:eps_crit}, we plot the critical confidence bound $\epsilon^*$ as a function of $M$.

\subsection{{Evidence} of Opinion Drift}\label{subsec:numerical_drift}

The ``drift'' of opinions refers to the unidirectional movement of a cluster of agents in opinion space. In the real world, one can interpret such drift behavior as a group of like-minded individuals simultaneously changing their opinions by following the same general trend. {A well-known example of opinion drift is herding behavior in financial markets, with investors making decisions based on the actions of other investors. This collective tendency can amplify
price movements and lead to significant mispricing~\cite{Banerjee1992}}. 

{Researchers have examined opinion drift in a variety of ABMs.}
{In BCMs, opinion drift} was first observed in the HK model with an asymmetric confidence bound, which preferentially allows compromises with a larger set of opinions on one side of an opinion~\cite{hegselmann2002opinion}.
Using a relative-agreement (RA) opinion model, Amblard and Deffuant~\cite{Amblard_Deffuant_2004} observed that opinions drift towards a single extreme on a specific family of small-world networks. 
{Stokes et al.~\cite{stokes2024} studied a BCM with interactions that depend on agents' historical opinion differences, and they observed that the mean opinion drifts for intermediate values of the confidence bound.}
{Sood} et al.~\cite{sood2008drift} examined opinion drift in biased voter models (VMs) on heterogeneous networks with large degree disparities.
In a recent study~\cite{Ramirez2024NonlinearVoter} of a nonlinear VM with finitely many opinion states, Ramírez et al.
 observed that, depending on the parameter regime, opinions can drift either toward more popular opinions or toward less popular opinions.
 {Ara\'{u}jo~\cite{araujo2026} studied an opinion model in which agents respond both to direct network interactions and to summaries of a
 population's opinion distribution, {and he demonstrated analytically that biased summaries can induce mean opinion drift {in polarized situations,}
 with {two opinion} clusters translating in the same direction while maintaining {a} constant separation at a critical parameter value.}

In our ABM, we observe opinion drift when $\epsilon > \epsilon^*$, which we recall results in a consensus state. In this parameter region, when the media opinion $M$ is sufficiently large, the dominant cluster moves (i.e., drifts) toward one of the media agents. This drift eventually stops, and the dominant cluster stabilizes at a fixed opinion value. In some cases, after the opinion drift stops, a second cluster can emerge on the opposite side of opinion space.
We illustrate such a scenario in Figure~\ref{fig:drift}, which shows 
the drift process and the formation of a second major opinion cluster.

\begin{figure}[p]
\centering
\resizebox{\textwidth}{!}{%
\begin{minipage}{\textwidth}

\begin{subfigure}{0.48\textwidth}
    \includegraphics[width=\linewidth]{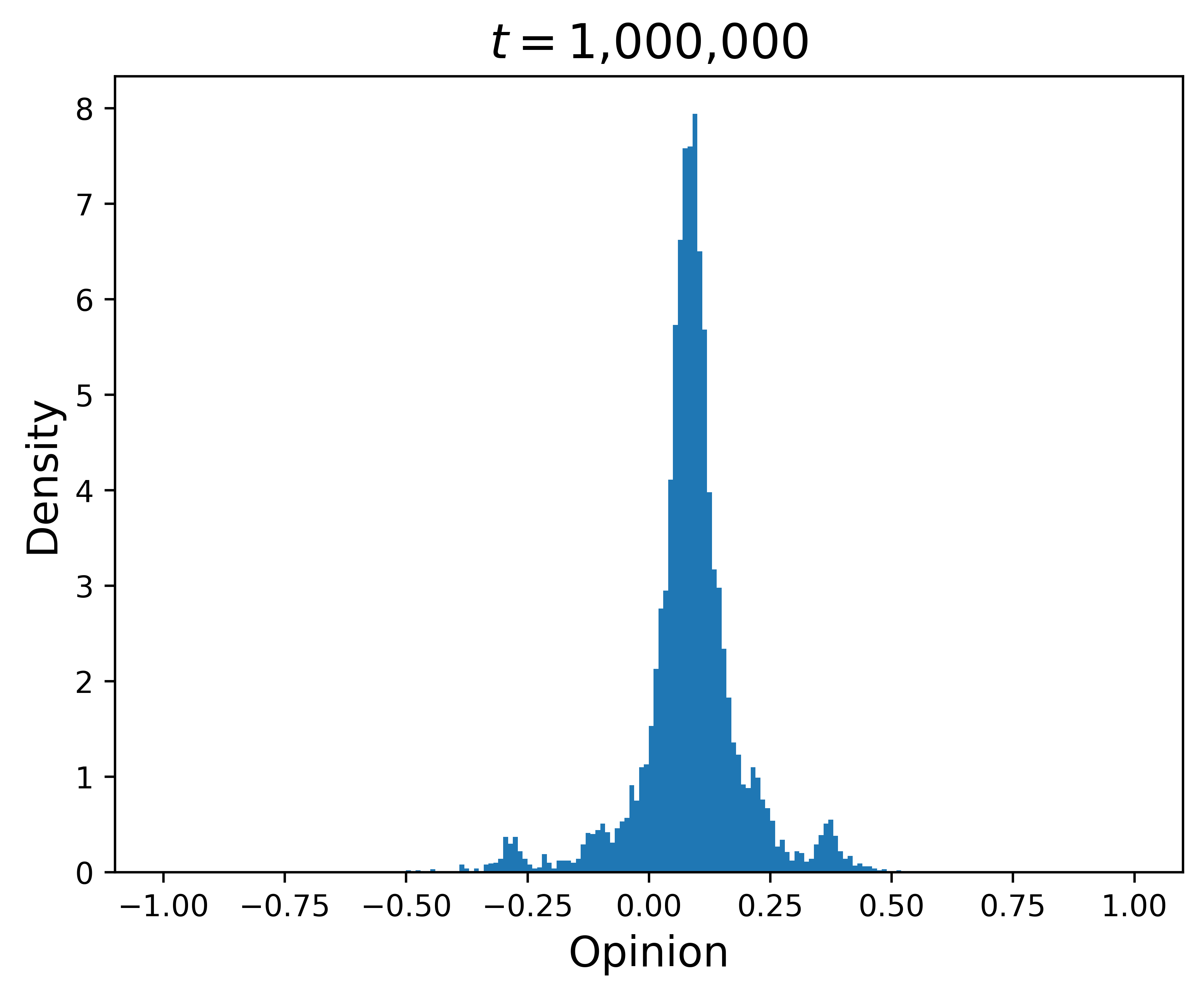}
\end{subfigure}\hfill
\begin{subfigure}{0.48\textwidth}
    \includegraphics[width=\linewidth]{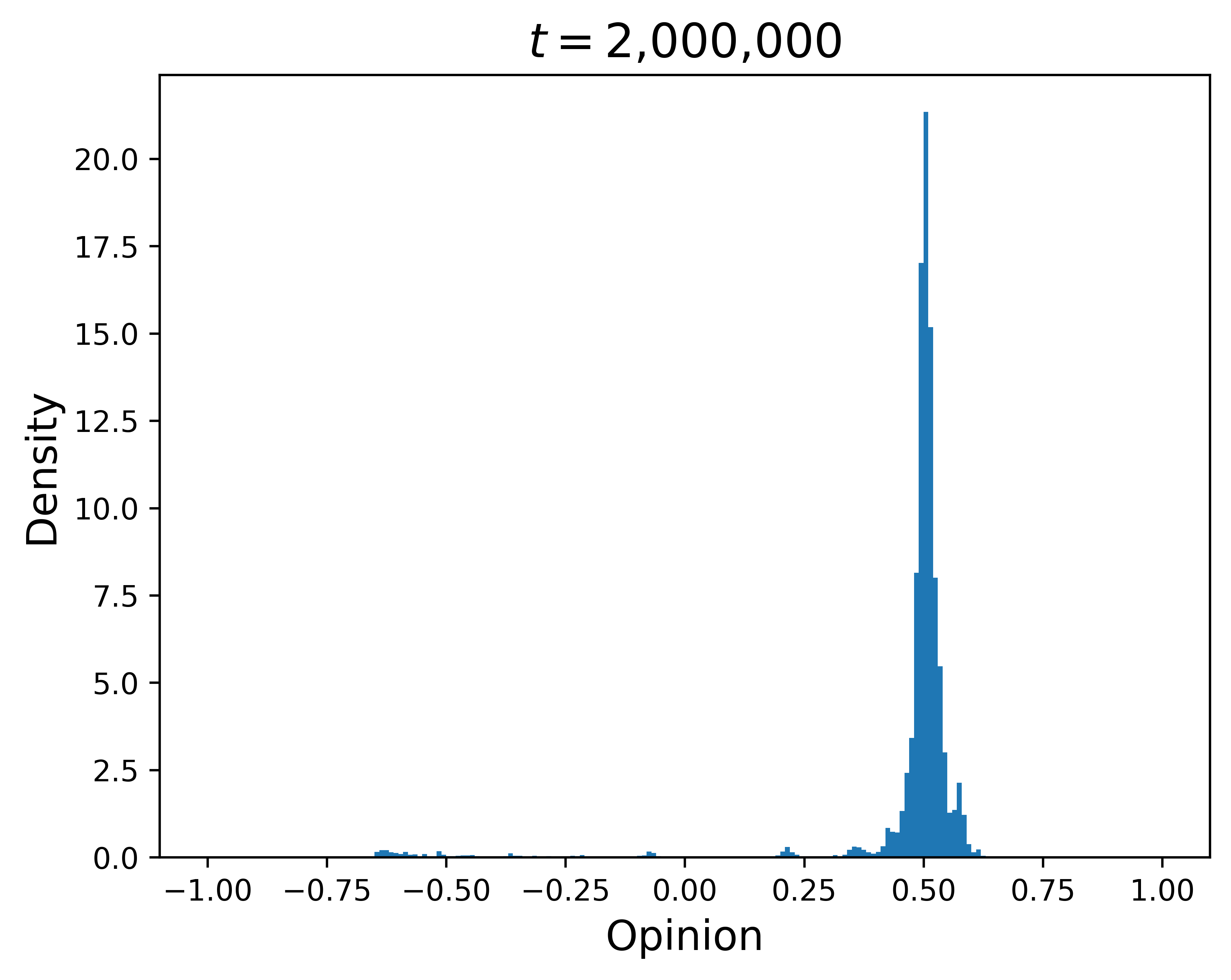}
\end{subfigure}

\vspace{0.5em}

\begin{subfigure}{0.48\textwidth}
    \includegraphics[width=\linewidth]{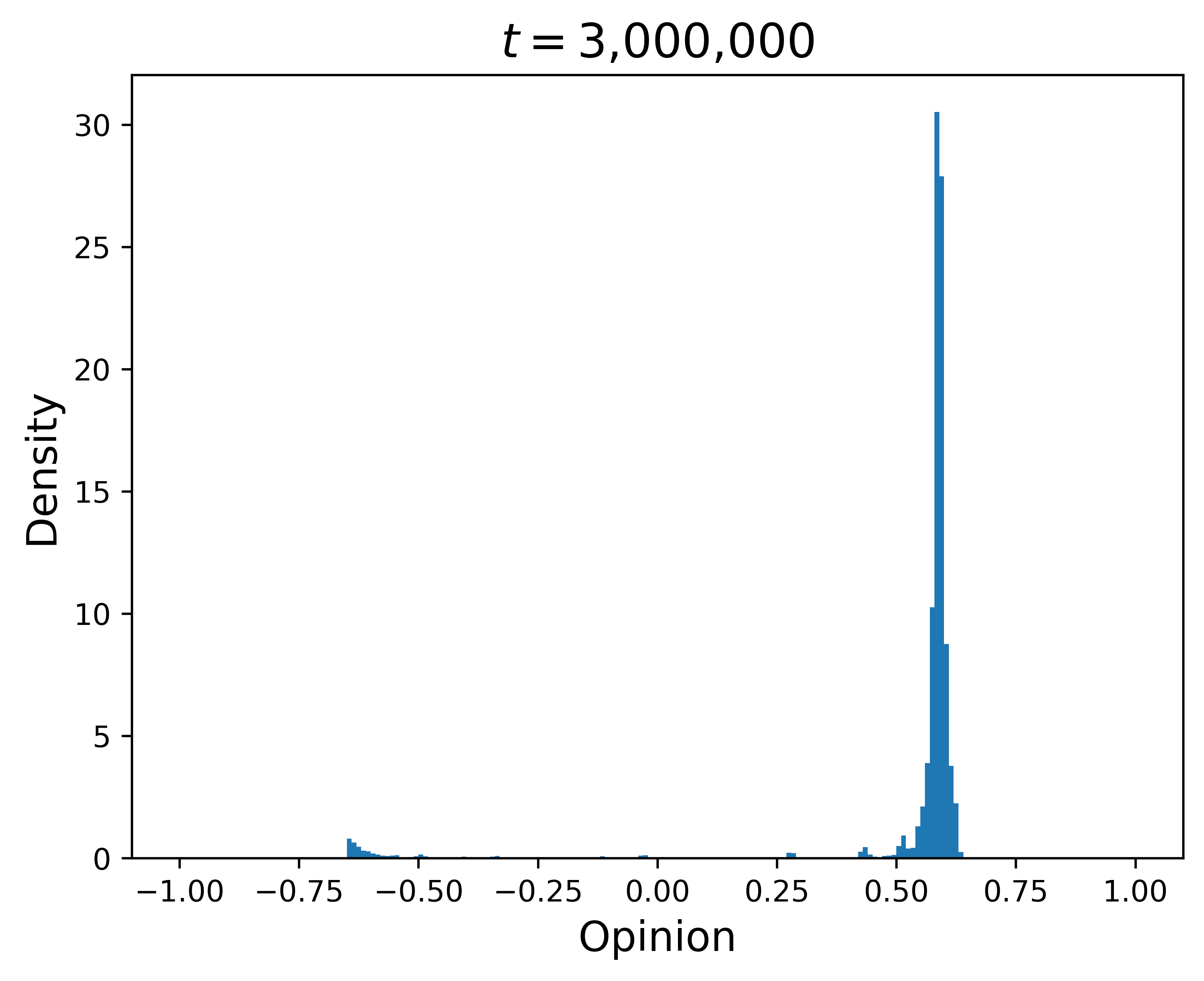}
\end{subfigure}\hfill
\begin{subfigure}{0.48\textwidth}
    \includegraphics[width=\linewidth]{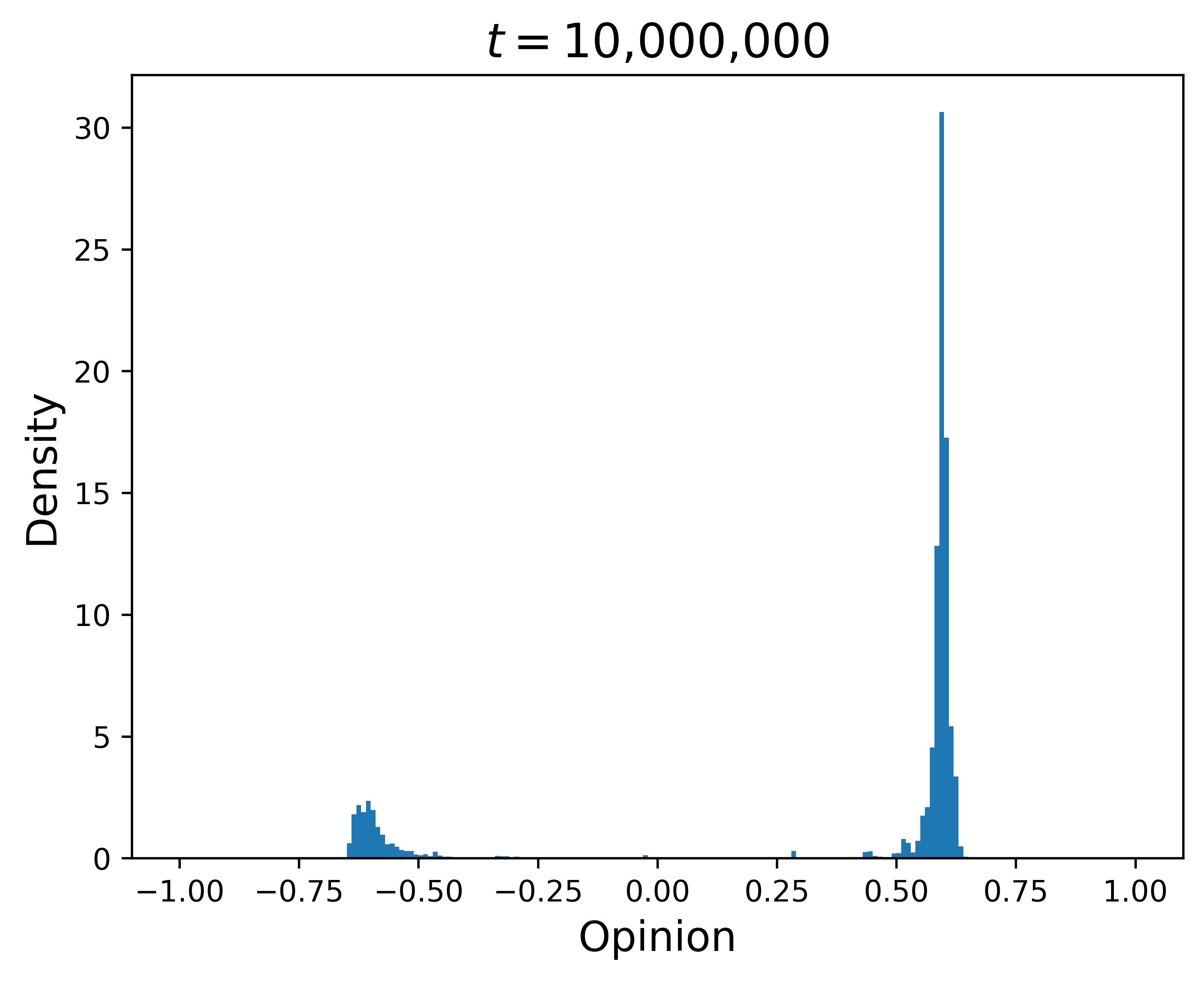}
\end{subfigure}

\vspace{0.5em}

\begin{subfigure}{0.48\textwidth}
    \includegraphics[width=\linewidth]{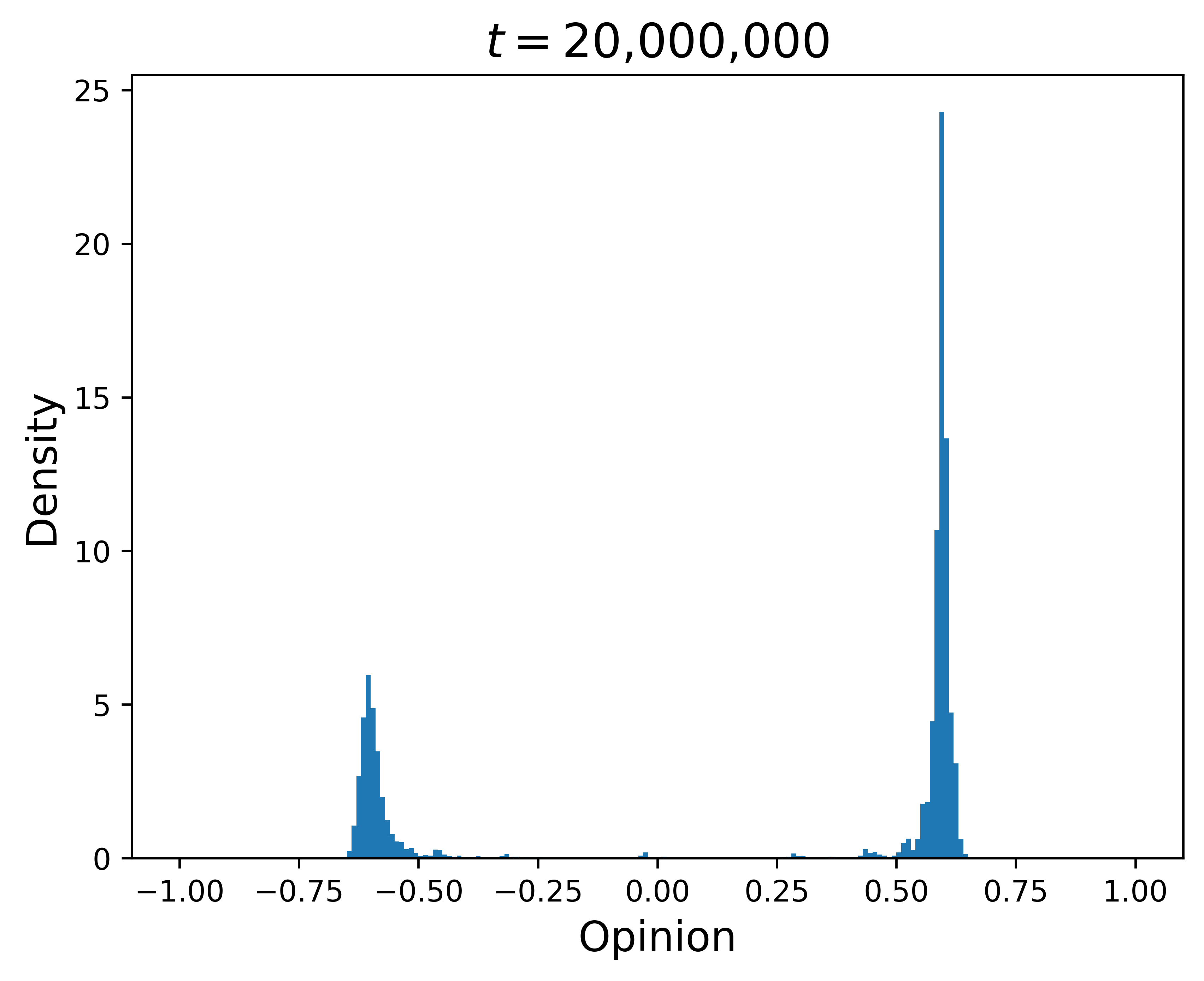}
\end{subfigure}\hfill
\begin{subfigure}{0.48\textwidth}
    \includegraphics[width=\linewidth]{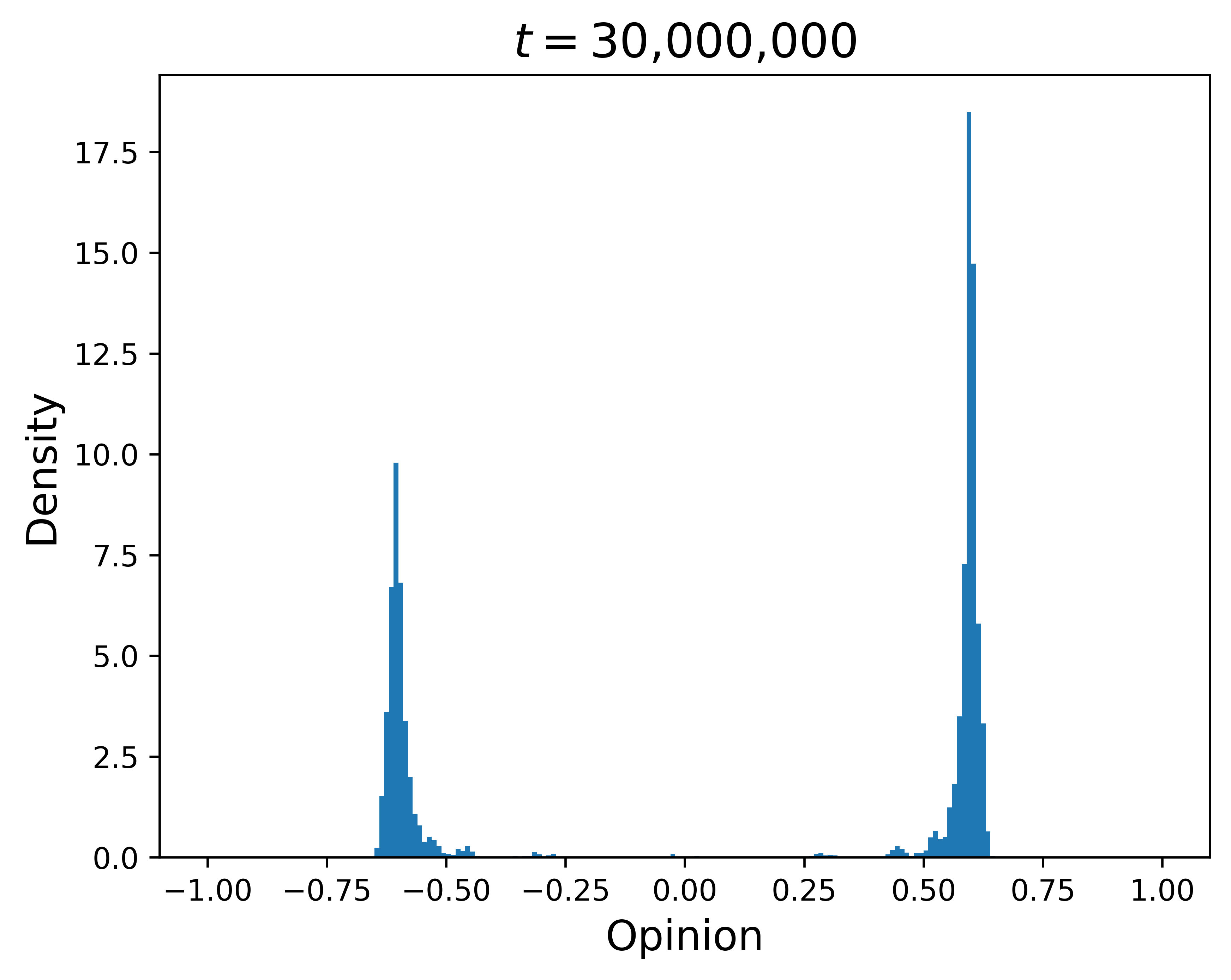}
\end{subfigure}

\vspace{0.5em}

\begin{subfigure}{0.48\textwidth}
    \includegraphics[width=\linewidth]{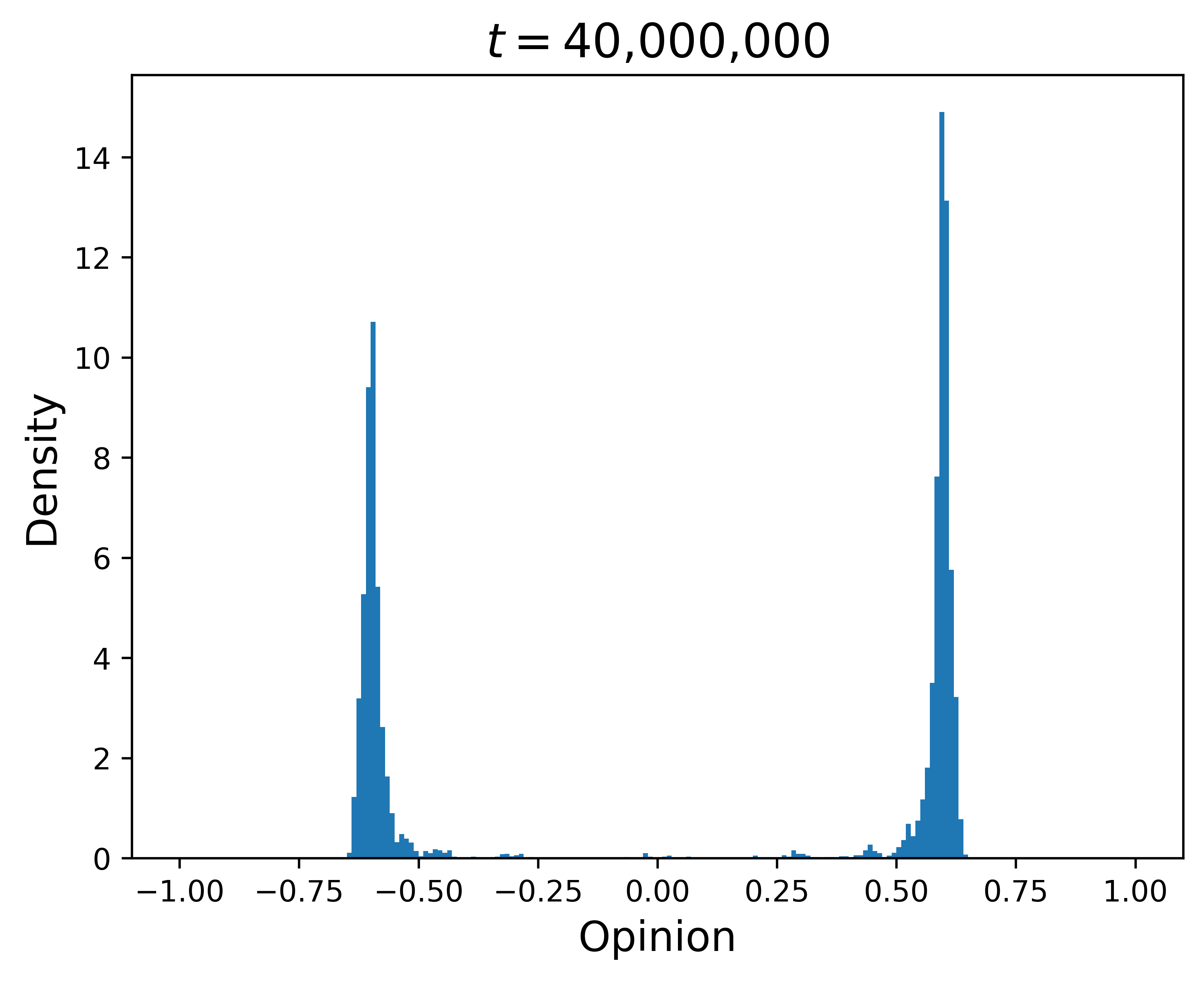}
\end{subfigure}\hfill
\begin{subfigure}{0.48\textwidth}
    \includegraphics[width=\linewidth]{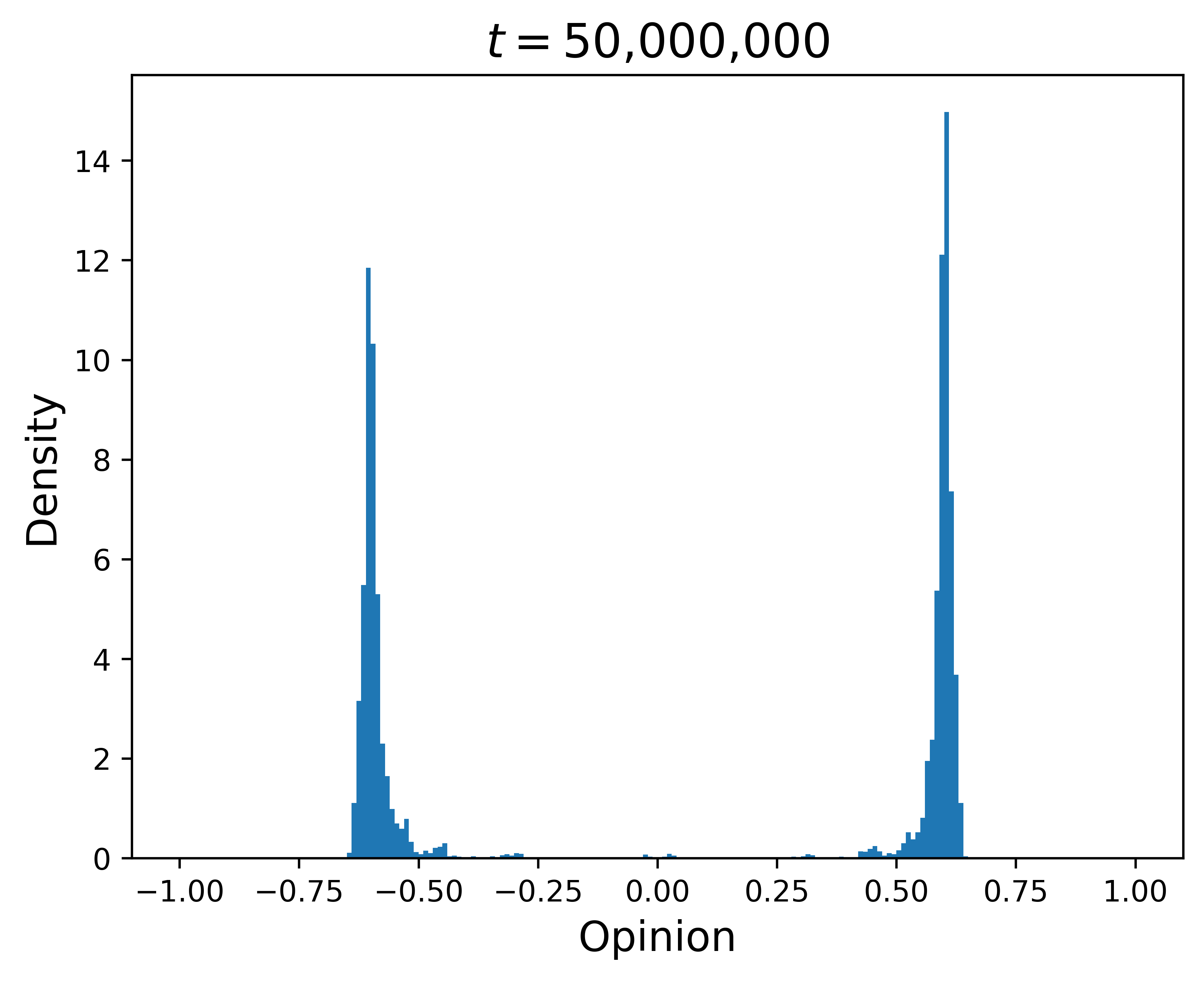}
\end{subfigure}

\end{minipage}
}
\caption{Opinion drift in one simulation of our ABM with confidence bound $\epsilon = 0.65$ and media opinion $M = 0.65$. The dominant opinion cluster drifts toward the positive media agent and stops at about time $t = \num{3000000}$. 
At about time $t = \num{10000000}$, a second major opinion cluster emerges
on the opposite side of opinion space. The second major opinion cluster grows until we halt the simulation at time $t = \num{50000000}$.
}
\label{fig:drift}
\end{figure}

We refer to the absolute value of the opinion where the dominant cluster stops drifting as a ``stable opinion'', which we denote by $x^*$. (We take the absolute value because the opinion cluster can drift to either the positive side or the negative side of opinion space.) If the dominant opinion cluster does not drift, we set $x^* = 0$. Based on our simulations, the stable opinion seems to be a function only of the media opinion $M$ and does not depend on the confident bound 
$\epsilon$. In Figure~\ref{fig:stable_points}, we show that opinion drift does not occur for small values of $M$ and only becomes apparent for $M \gtrapprox 0.4$.

\begin{figure}[htbp]
    \centering
    \includegraphics[width=0.8\textwidth]{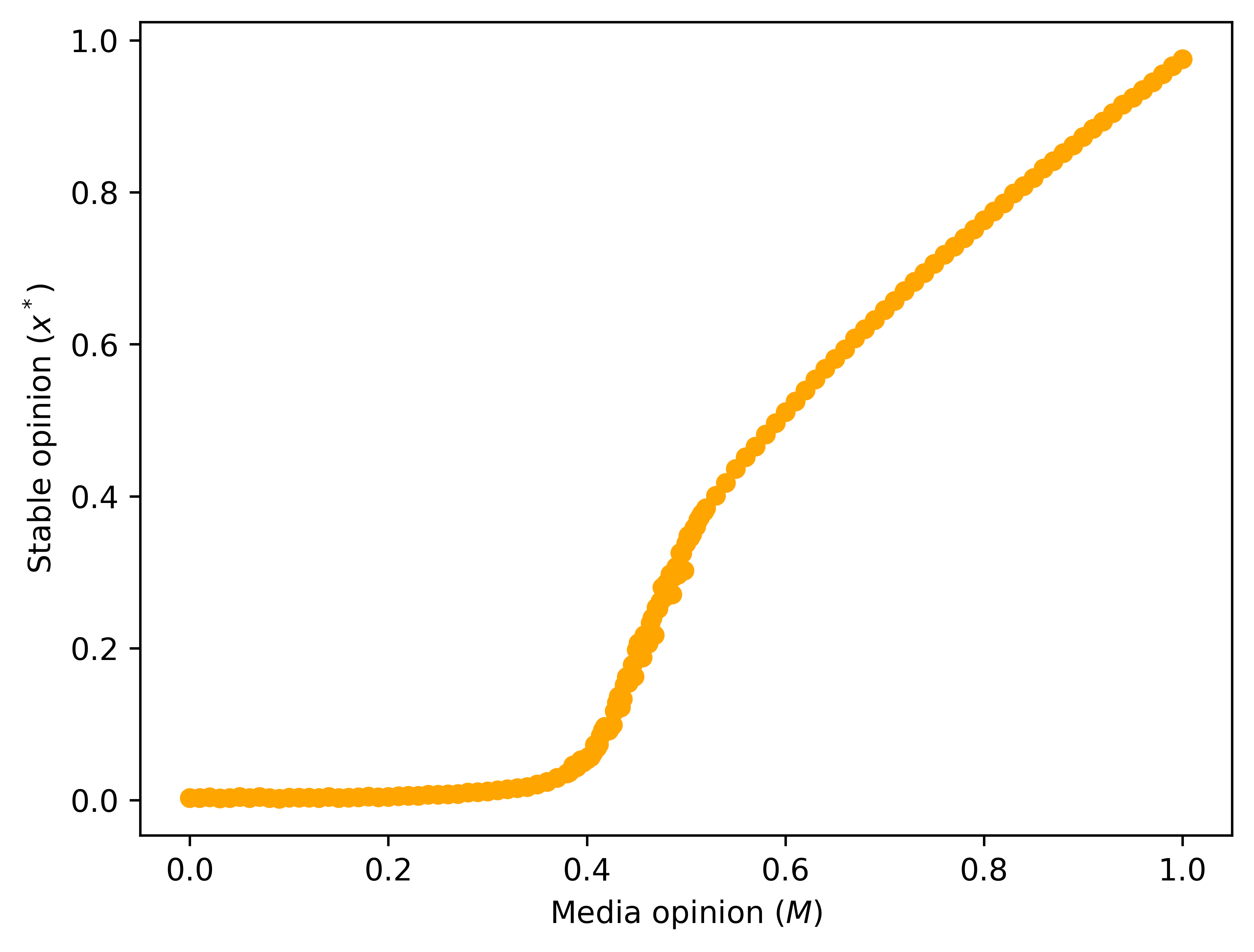}
    \caption{Absolute value of the opinion where the dominant opinion cluster stops drifting as a function of the media opinion $M$. Each data point represents the result of one simulation, and we stop each simulation after $\num{5000000}$ time steps. {We stop these simulations after $\num{5000000}$ time steps rather than after $\num{50000000}$ time steps as in Fig.~\ref{fig:drift} because we care only about the drift of the dominant opinion cluster (and not the subsequent emergence of a second cluster).} We take the maximum absolute value of the mean of the opinions of the dominant cluster's constituent nodes to be the stable opinion.
    }
    \label{fig:stable_points}
\end{figure}

\begin{figure}[htbp]
    \centering
    \includegraphics[width=0.8\textwidth]{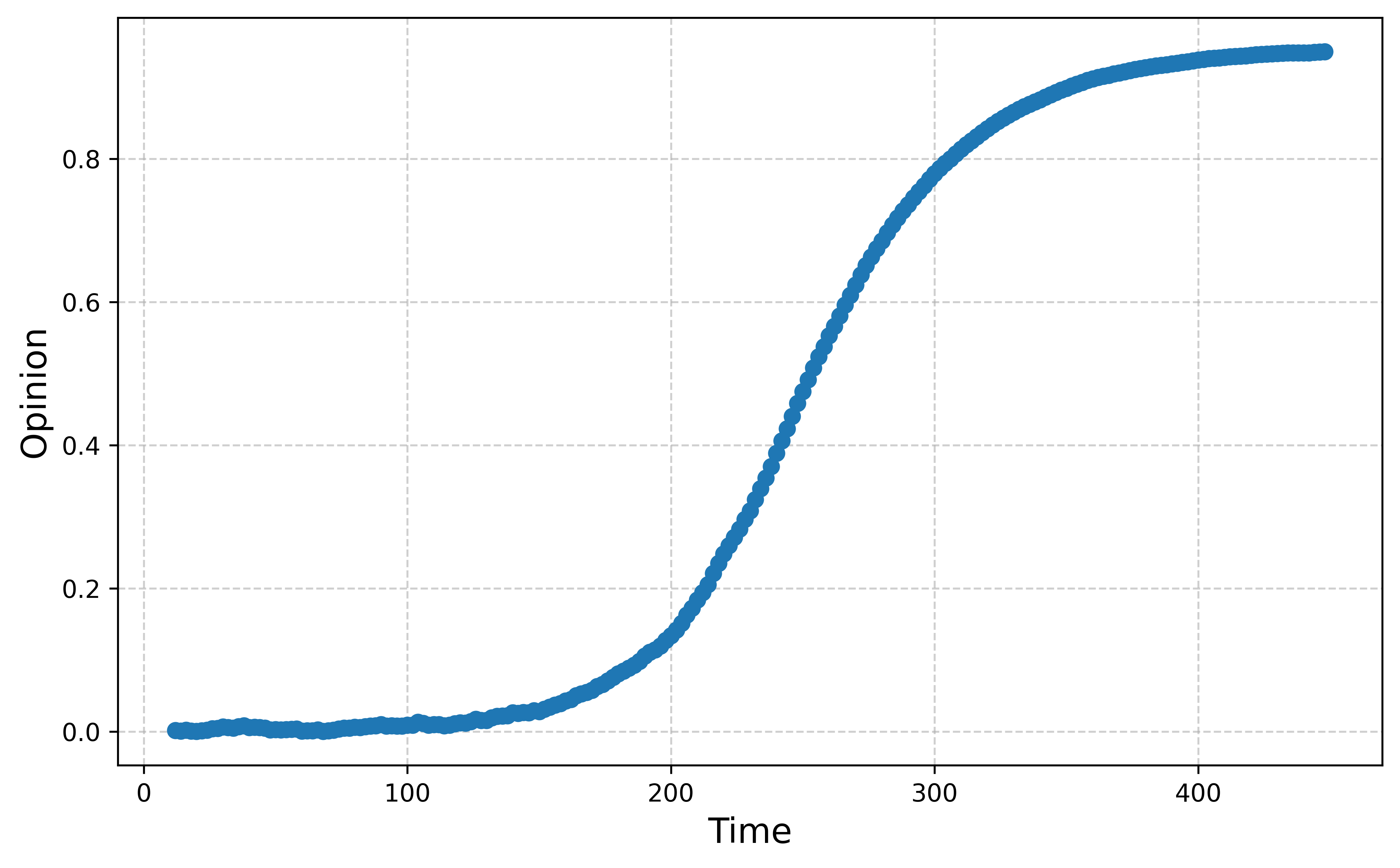}
    \caption{Location of the dominant opinion cluster as a function of time in one simulation when the media opinion is $M = 1$. We take the mean of the opinions of the constituent nodes to be the opinion of the dominant cluster. We stop the simulation after $\num{2500000}$ time steps.}
    \label{fig:drift_trajectory}
\end{figure}

Using our simulations, we also plot the ``trajectory'' of the dominant opinion cluster's drift (see Figure~\ref{fig:drift_trajectory}), which also appears to be independent of the confidence bound $\epsilon$. It resembles a logistic curve, and 
it {approaches} $x^*(M)$ in the long-time limit.

In Section~\ref{sec:mfa}, we derive {continuous-time} equations for our ABM. In Section~\ref{sec:drift}, we use these equations to approximate the stable opinion and drift trajectory as functions of the media opinion $M$.


\section{{Continuous-Time} Approximation}\label{sec:mfa}

To better understand the mechanisms that drive the behavior of our ABM, we derive a {continuous-time} approximation of it and thereby obtain an ordinary differential equation that approximately describes its mean behavior.

Ben-Naim, Krapivsky, and Redner~\cite{bennaim2003bifurcations} used a similar continuous-time approach to study the DW model in a mean-field setting, and such approaches subsequently have been used by many other researchers~\cite{chu2024inference,dubovskaya2023analysis,fennell2021generalized}.

To obtain the desired {continuous-time} approximation, we first change from discrete time to continuous time.
Let $\Delta t$ denote the duration of one discrete time step in the model, and recall that $N$ denotes the number of nodes in a network.
One unit of continuous time is $\frac{N\Delta t}{2}$; {during a single time step, each agent has
approximately one interaction on average.

{To describe the expected change in the opinion of an agent $i$ in one time step, we write}
\begin{equation}\label{eq:change in x_i}
	x_i(t + \Delta t) - x_i(t) = (1 - p_M) I_1 + p_M I_2 \, ,
\end{equation} 
where the term $I_1$ arises from interactions between agent $i$ and other ordinary agents and the term $I_2$ arises from interactions between agent $i$ and media agents.

{To obtain $I_1$, we examine the expected change in agent $i$'s opinion} $x_i$ due to its interactions with ordinary agents $j$. 
The probability of selecting the pair $(i, j)$ uniformly at random is
 $\frac{2}{N(N - 1)}$, and agents $i$ and $j$
 compromise if and only if} $|x_i - x_j| < \epsilon$. If they compromise, $x_i$ increases by $\frac{x_j - x_i}{2}$. Summing over all ordinary
 agents $j$ then yields
\begin{equation}\label{eq:i1}
    I_1 = \sum_{j = 1}^N \frac{2}{N(N - 1)} \cdot u(\epsilon - |x_i - x_j|) \left(\frac{x_j - x_i}{2}\right) \, ,
\end{equation} 
{where $u(\cdot)$ is the step function} 
\begin{equation*}
	u(x) =
		\begin{cases}
		    1 & \text{if } x \ge 0 \\
		    0 & \text{if } x < 0 \, . 
		\end{cases}
\end{equation*} 

{To obtain $I_2$, we examine the expected change in $x_i$ due to its interactions with the two media agents. The probability of selecting agent $i$ is ${1}/{N}$, and the probability that it interacts with one of the media agents is $p_M$.
If agent $i$ interacts with the positive media agent, which occurs with probability $p_+(x_i)$, then its opinion $x_i$ changes by $\frac{M - x_i}{2}$ (and hence increases). If agent $i$ instead interacts with the negative media agent, which occurs with probability $p_-(x_i)$, then its opinion $x_i$ changes by $\frac{-M - x_i}{2}$ (and hence decreases). Summing the two associated terms yields}
\begin{equation}\label{eq:i2}
    I_2 = \frac{1}{N} \left[p_+(x_i)\left(\frac{M - x_i}{2}\right) + p_-(x_i) \left(\frac{-M - x_i}{2}\right)\right] \, .
\end{equation} 

Substituting~\eqref{eq:i1} and~\eqref{eq:i2} into~\eqref{eq:change in x_i} and dividing both sides by $\Delta t = {2}/{N}$ yields
\begin{align*}
    \frac{x_i(t + \Delta t) - x_i(t)}{\Delta t} &= (1 - p_M) \frac{1}{N - 1} \sum_{j = 1}^N u(\epsilon - |x_i - x_j|) \left(\frac{x_i - x_j}{2}\right) \\
    			&\; \; \; \; + \frac{1}{2}p_M\left[p_+(x_i)\left(\frac{M - x_i}{2}\right) + p_-(x_i)\left(\frac{-M - x_i}{2}\right)\right] \,,
\end{align*} 
and taking the limit $N \rightarrow \infty$ then gives
\begin{equation}
	\begin{aligned}
		\frac{\mathrm{d}x_i}{\mathrm{d}t} &= (1 - p_M) \lim_{N\rightarrow \infty} \frac{1}{N}\sum_{j = 1}^{N}u(\epsilon - |x_i - x_j|) \left(\frac{x_j - x_i}{2}\right) \\
			&\quad +   \frac{1}{2}p_M\left[p_+(x_i) \left(\frac{M - x_i}{2}\right) + p_-(x_i) \left(\frac{-M - x_i}{2}\right)\right]\, .
	\end{aligned}
\label{eq:cts_time}
\end{equation} 

The continuous-time approximation~\eqref{eq:cts_time} is of particular importance to us in Section~\ref{sec:drift} for deriving analytical approximations of the opinion drift.


\section{Opinion Drift}\label{sec:drift}

We now use the continuous-time BCM~\eqref{eq:cts_time} to analytically study the drift phenomenon that we observed in Section~\ref{subsec:numerical_drift}. To simplify our analysis, we assume without loss of generality that the dominant opinion cluster always drifts toward the positive media agent. We also assume that the dominant opinion cluster behaves as a single unit with location $x_c(t)$ and that no agents have opinions that lie outside this cluster. Under these assumptions, we can neglect interactions between ordinary agents and we can write an ordinary differential equation in terms of $x_c(t)$. We thereby obtain

\begin{equation}
	\begin{aligned}
		\frac{\mathrm{d}x_c}{\mathrm{d}t} &= \frac{1}{2}p_M\left[p_+(x_c) \left(\frac{M - x_c}{2}\right)
              +   p_-(x_c) \left(\frac{-M - x_c}{2}\right)
           \right] \\
           &=  \frac{1}{4}p_M\left[\frac{1}{1  +  e^{-5x_c}}\cdot(M - x_c) + \frac{1}{1  +  e^{5x_c}}\cdot(-M - x_c)\right] \\
           &= \frac{1}{4}p_M\left[M\tanh\left(\frac{5x_c}{2}\right) - x_c\right] \, .
	\end{aligned}
\label{eq:drift}
\end{equation}


\subsection{Stable Opinion} \label{subsec:stable_points}

In Section~\ref{subsec:numerical_drift}, we defined the stable opinion $x^*$ as the opinion value where the dominant opinion cluster stops drifting. (If it does not drift at all, we set $x^* = 0$.) In the present subsection, we consider nonnegative stable opinions. A stable opinion must satisfy
\begin{align*}
	\frac{\mathrm{d}x_c}{\mathrm{d}t} = 0 \quad \implies \quad \frac{1}{4}p_M\left[M\tanh\left(\frac{5x^*}{2}\right) - x^*\right] = 0 \, ,
\end{align*}
which implies that
\begin{align}
	x^* = M\tanh\left(\frac{5x^*}{2}\right) \, .
\label{eq:stable_points}
\end{align}
Note that $x^* = 0$ is a trivial solution of Eq.~\eqref{eq:stable_points}.

\begin{proposition} \label{prop:stable_points}
When the media opinion $M \le 0.4$, the only solution of Eq.~\eqref{eq:stable_points}
is the trivial solution $x^* = 0$.
\end{proposition}

\begin{proof}
Let 
\begin{equation}
\label{eq:f(x)}
	f(x) = M\tanh\left(\frac{5x}{2}\right) - x \, ,
\end{equation}
which implies that
\[
	f'(x) = M\cdot \frac{5}{2}\sech^2\left(\frac{5x}{2}\right) - 1 \, .
\]
Because $0 < \sech^2(x) \le 1$ for all $x$, we have
\[
	f'(x)  \le  \frac{5M}{2} - 1 \, .
\]
If $\frac{5M}{2} - 1 < 0$ (i.e., $M < 0.4$), then $f'(x) < 0$ for all $x$, so $f$ is strictly decreasing. A strictly decreasing continuous function has at most one zero. Because $f(0) = 0$, the unique solution is $x = 0$.
\end{proof}


\subsection{Drift Region: $M > 0.4$}

Proposition~\ref{prop:stable_points} implies that only the trivial stable opinion exists for $M \leq 0.4$ and hence that there is no opinion drift. When $M > 0.4$, nonzero stable opinions appear and there is opinion drift. From Eq.~\eqref{eq:stable_points}, these stable opinions are 
\begin{equation}
	x^* = \frac{4}{5}\operatorname{cosh}^{-1}\left(\sqrt{\frac{5M}{2}}\right) \, .
\label{eq:stable_points2}
\end{equation}

Combining Proposition~\ref{prop:stable_points} and Eq.~\eqref{eq:stable_points2}, we conclude that the largest stable opinion as a function of $M$ is
\begin{equation}
	x^*(M) = 
		\begin{cases}
		    0 & \text{if } M \le 0.4 
		    \\
		    \frac{4}{5}\operatorname{cosh}^{-1}\left(\sqrt{\frac{5M}{2}}\right) & \text{if } M > 0.4 \, .
\end{cases}
\label{eq:stable_points3}
\end{equation}

In Figure~\ref{fig:stable_points_comp}, we compare Eq.~\eqref{eq:stable_points3} with {the results of a numerical simulation of our ABM}.

\begin{figure}[htbp]
    \centering
    \includegraphics[width=0.8\textwidth]{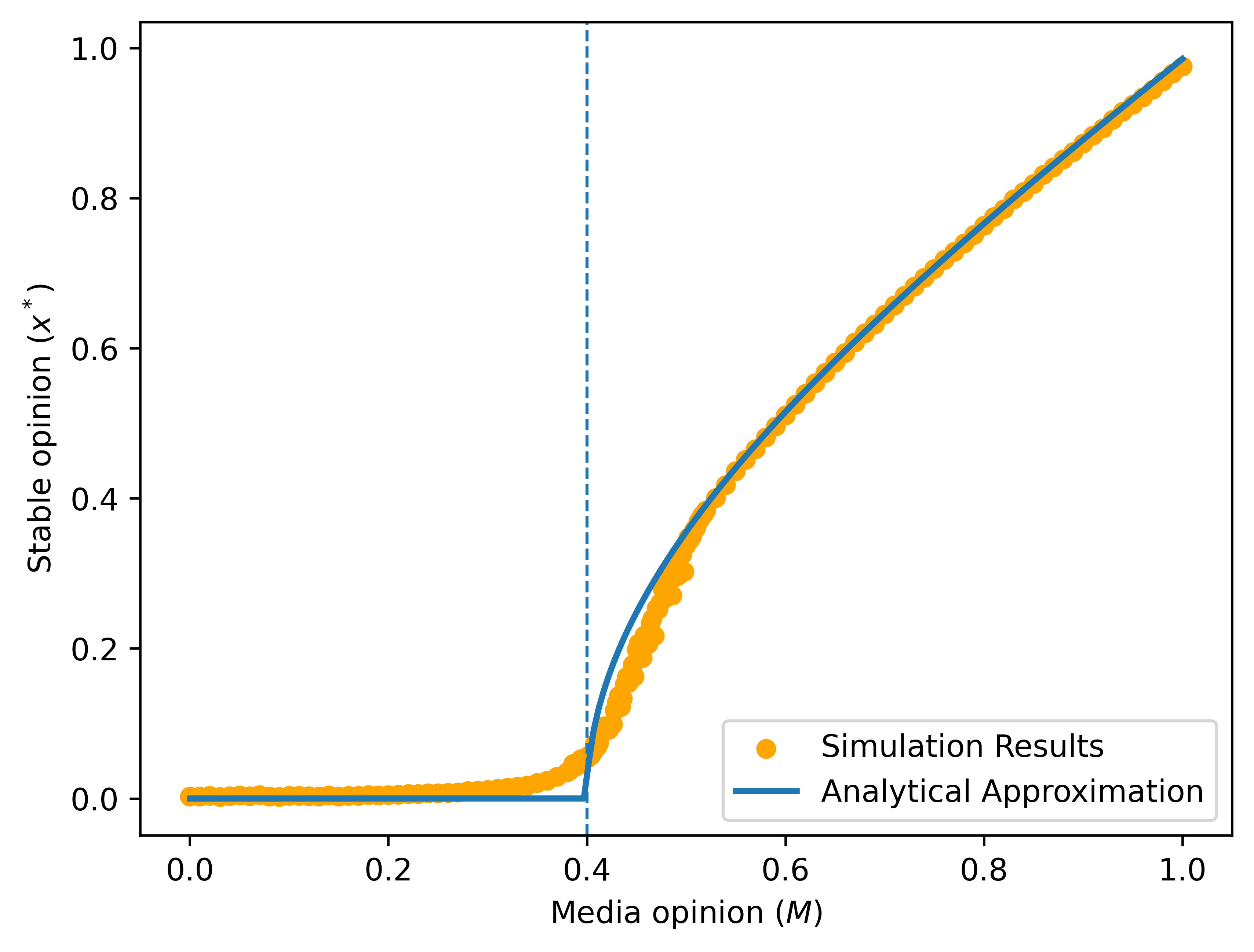}
    \caption{Stable opinions that we obtain numerically overlaid with our analytical solution \eqref{eq:stable_points3} for them. To obtain the numerical results, we conduct one simulation of our ABM for each media opinion 
    between $M = 0$ and $M = 1$ in increments of $0.01$. The dashed blue line indicates the media opinion $M = 0.4$. We plot the largest stable opinion for each value of $M$.
    }
    \label{fig:stable_points_comp}
\end{figure}

We aim to analytically describe the drift trajectory, which we illustrated in Figure~\ref{fig:drift_trajectory}. The trajectory resembles a logistic function, {which is a solution of the logistic equation} 
\begin{equation}\label{eq:logistic_diffeq}
    \frac{dx}{dt} = kx\left(1 - \frac{x}{L}\right) \, .
\end{equation} 
{Because the right-hand side of~\eqref{eq:logistic_diffeq} depends quadratically on} $x$, we expect the right-hand side of \eqref{eq:drift} to be approximately quadratic.
Indeed, as we see in Figure~\ref{fig:fitted_quadratic}, $f(x_c) = \frac{\mathrm{d}x_c}{\mathrm{d}t}$ closely resembles a quadratic function of $x_c$ in the region $[0, M]$.
To formalize this idea, we assume that $f(x_c)$ is approximately of the form
\begin{equation}
	f(x_c) \approx kx_c\left(1 - \frac{x_c}{L}\right) \, ,
\label{eq:logistic}
\end{equation} 
where $L$ is the asymptotic value of $x_c$ as time $t \rightarrow \infty$.
 In our ABM, the asymptotic value is a stable opinion, so
\begin{equation}
	L = x^*(M) = \frac{4}{5}\operatorname{cosh}^{-1}\left(\sqrt{\frac{5M}{2}}\right) \, .
\label{eq:asymptote}
\end{equation} 
We determine the coefficient $k$ by equating local maxima of $f(x_c)$ and the quadratic~\eqref{eq:logistic}. Setting $f'(x_c)=0$ gives
\begin{equation*}
	\frac{5M}{2} \operatorname{sech}^{2}\left(\frac{5x_c}{2}\right) - 1 = 0 \quad \implies \quad x_c = \frac{2}{5}\operatorname{sech}^{-1}\left(\sqrt{\frac{2}{5M}}\right) \, .
\end{equation*} 
The corresponding maximum of $f(x_c)$ is
\begin{equation*}
	M\sqrt{1 - \frac{2}{5M}} - \frac{2}{5}\operatorname{sech}^{-1}\left(\sqrt{\frac{2}{5M}}\right) \, .
\end{equation*} 
Equating this maximum to the maximum of the quadratic gives
\begin{equation*}
	M\sqrt{1 - \frac{2}{5M}} - \frac{2}{5}\operatorname{sech}^{-1}\left(\sqrt{\frac{2}{5M}}\right) =  \frac{kL}{4} =  \frac{k}{5}\operatorname{cosh}^{-1}\left(\sqrt{\frac{5M}{2}}\right) \, ,
\end{equation*}
which yields
\begin{equation}
	k = \frac{5M\sqrt{1-\frac{2}{5M}} - 2\operatorname{cosh}^{-1}\left(\sqrt{\frac{5M}{2}}\right)}{\operatorname{cosh}^{-1}\left(\sqrt{\frac{5M}{2}}\right)} \, .
\end{equation}

\begin{figure}[htbp]
    \centering
    \includegraphics[width=0.8\textwidth]{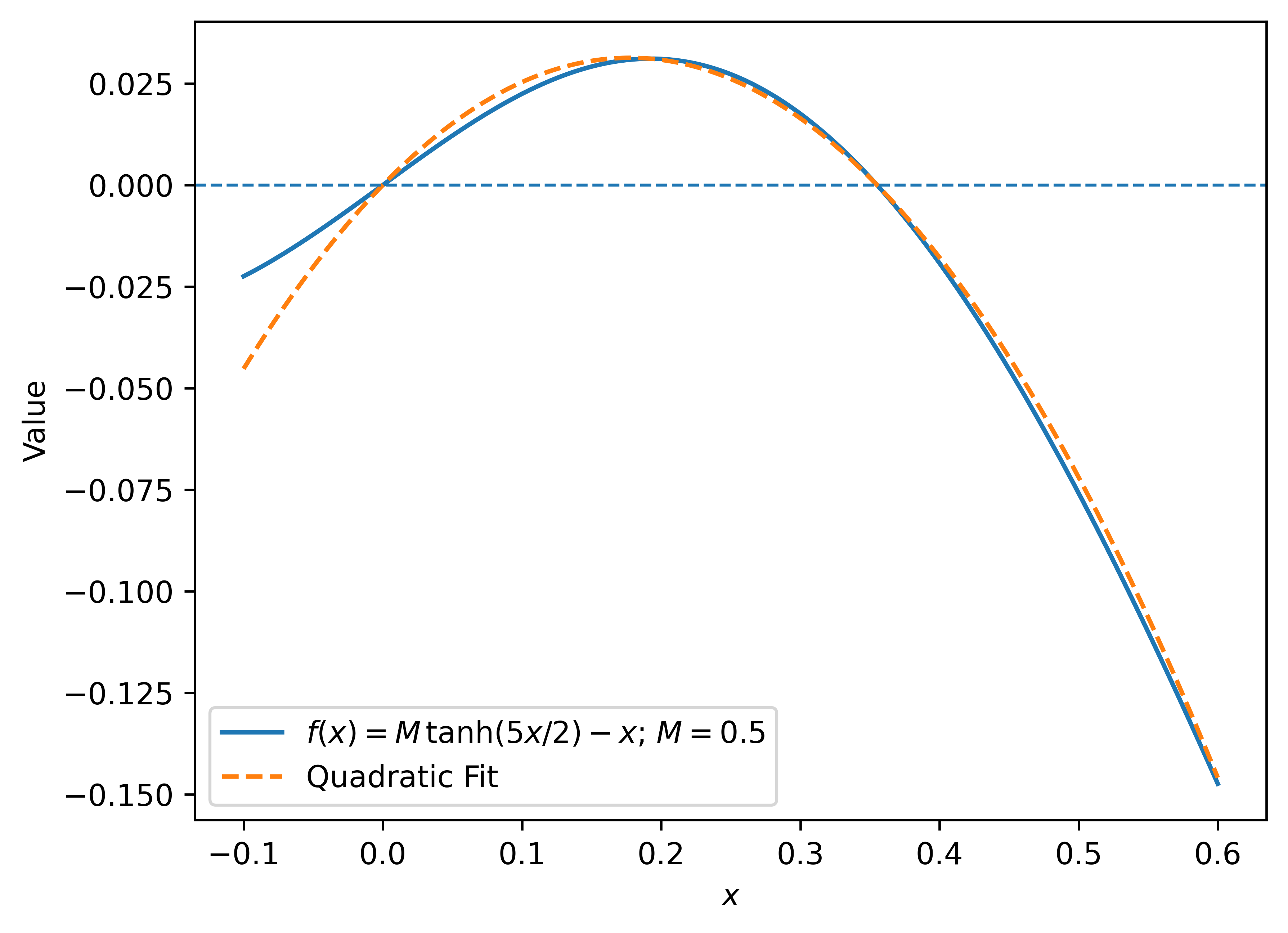}
    \caption{Comparison of the function $f(x)$ (see \eqref{eq:f(x)}) to a fit quadratic function.
    In this plot, $M = 0.5$ and we scale the quadratic to match the zeros and local maximum of $f(x)$.
    }
    \label{fig:fitted_quadratic}
\end{figure}

Our final equation for the predicted drift trajectory takes the form
\begin{equation}
	x_c(t) = \frac{L}{1 + Ae^{-kt}} \, .
\label{eq:drift_traj}
\end{equation} 
With the initial opinion-cluster location $x_c(0) = 0.01$, we have
\begin{equation*}
	A = 100L - 1 \, .
\end{equation*}
{In Figure~\ref{fig:drift_trajectory_comp}, we compare our analytical approximation \eqref{eq:drift_traj} with the results of a numerical simulation {of our ABM}.} 
There is a significant discrepancy between the simulated and analytically approximated 
{drift} trajectories, with the analytical approximation yielding a noticeably larger opinion-cluster location than the one from simulations 
after about $t = 250$.

One possible reason is that $N = \num{10000}$ is not a sufficiently large number of agents. We derive the
 continuous-time approximation~\eqref{eq:cts_time} for $N \to \infty$. As one can see in Figure~\ref{fig:drift_trajectory_comp_N}, there is a shift in the drift trajectory that depends on $N$, with 
 larger values of $N$ resulting in faster approaches to a stable opinion-cluster location.
  However, the stable opinion-cluster locations in these numerical simulations are still noticeably below the stable opinion-cluster location in our analytical approximation.
In Section~\ref{subsubsec:refinement}, we will attempt to resolve this discrepancy by refining our analytical approximation.

\begin{figure}[htbp]
    \centering
    \includegraphics[width=0.8\textwidth]{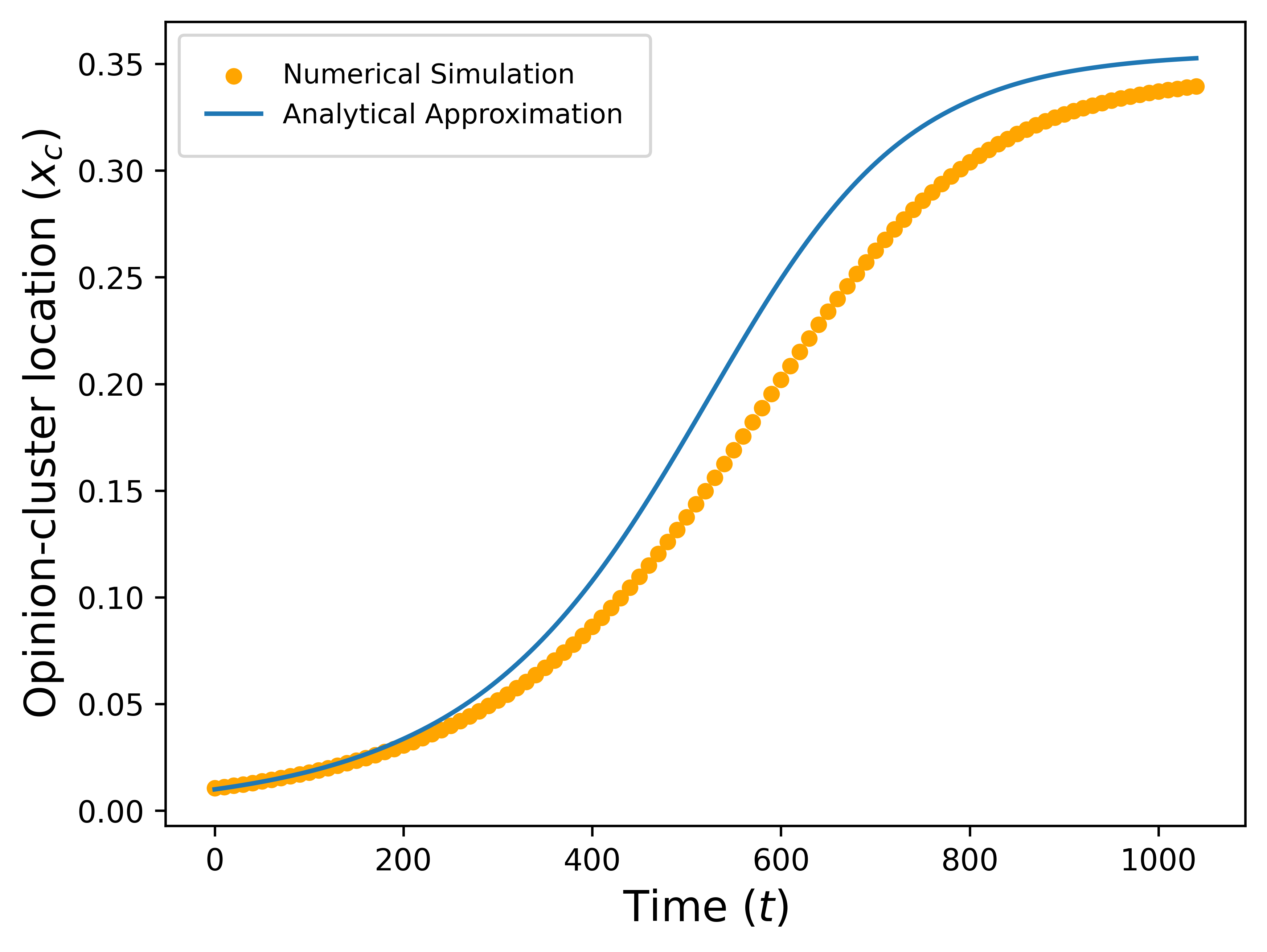}
    \caption{Comparison between our analytical approximation~\eqref{eq:drift_traj} of the drift trajectory and the result of a numerical simulation of our ABM.
    In this plot, the media opinion is $M = 0.5$ and the initial opinion-cluster location is $x_c(0) = 0.01$.
    }
    \label{fig:drift_trajectory_comp}
\end{figure}

\begin{figure}[htbp]
    \centering
    \includegraphics[width=0.8\textwidth]{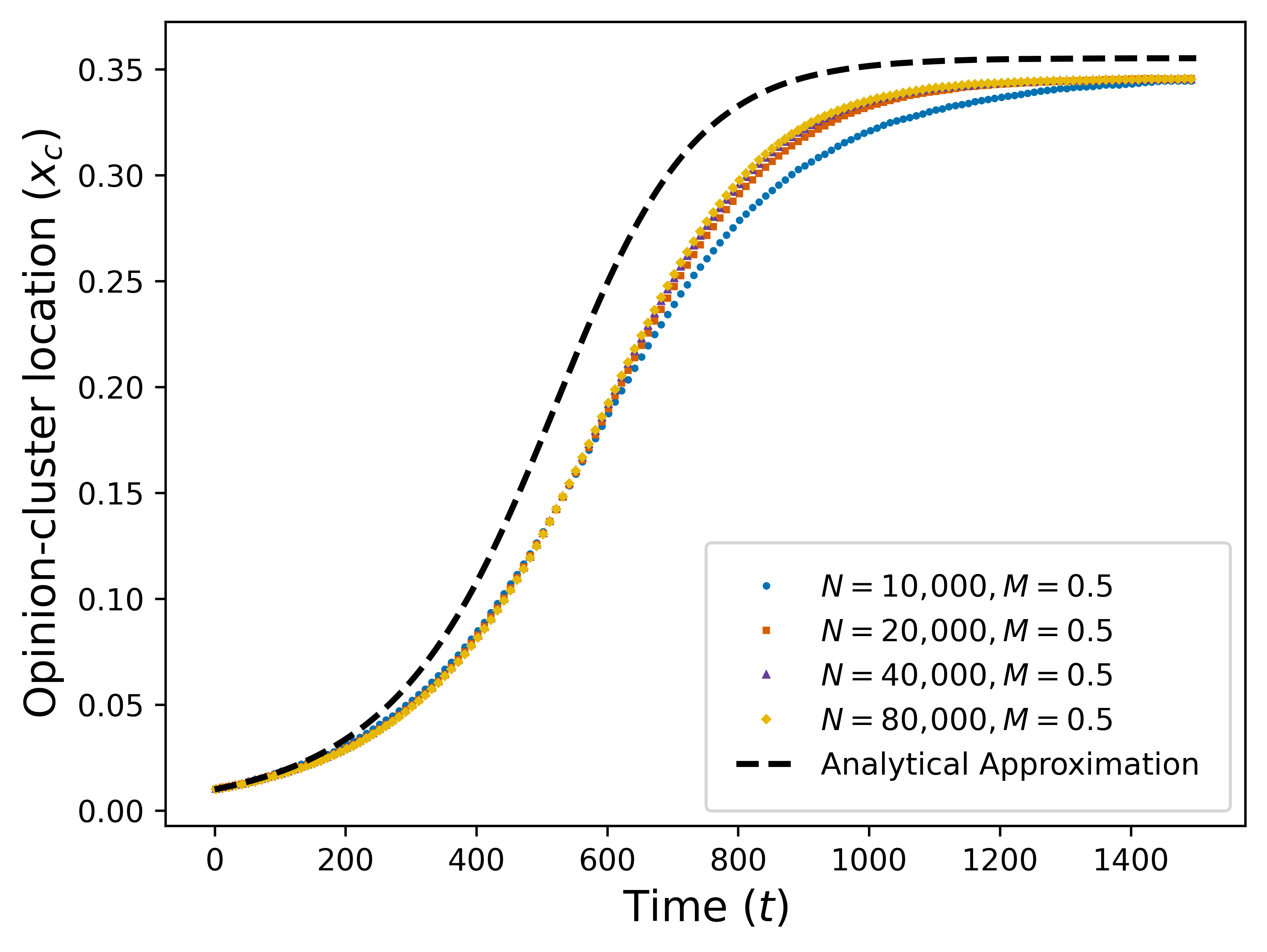}
    \caption{Comparison between our analytical approximation~\eqref{eq:drift_traj} of the drift trajectory and the results of numerical simulations of our ABM for $N = \num{10000}$, $N = \num{20000}$, $N = \num{40000}$, and 
    $N = \num{80000}$ agents. For each value of $N$, we run 100 simulations and took the mean. To avoid variations in initial conditions due to the value of $N$, all agents have initial opinions of $0.01$. In all simulations, 
    the media opinion is $M = 0.5$.
    }
    \label{fig:drift_trajectory_comp_N}
\end{figure}


\subsubsection{Drift Timescale}

To estimate the amount of time that the dominant opinion cluster drifts, we measure the time that it takes to reach the inflection point $x_c = x^*/2$ from $x_c = 0.01$, and we then double this time. Using~\eqref{eq:logistic}, we have
\begin{equation}
	x_c(t)  = \frac{L}{1 + e^{-kt}} \quad \implies \quad t(x_c) = -\frac{1}{k}\ln\left(\frac{L}{x_c} - 1\right) \, .
\end{equation}

The drift time $t_d$ is
\begin{equation}
	\begin{aligned}
		t_d(M)   &=  t\left(\frac{x^*}{2}\right) - t(0.01)  \\
				&=  \frac{2\ln\Big(100L(M) - 1\Big)}{C\cdot k(M)} \, ,
	\end{aligned}
\label{eq:drift_time}
\end{equation}
where
\begin{equation*}
	C  =  \frac{1}{2}\left(1 - e^{-\frac{p_M}{2}}\right) \,,
\end{equation*} 
where we emphasize the dependence on the media opinion $M$ by writing $k = k(M)$ and $L = L(M)$. In Figure~\ref{fig:drift_time_comp}, we compare this analytical estimate to results of numerical simulations 
{of our ABM}. We observe significance discrepancies between analytically estimated and simulated drift times for most $M \in [0.41, 0.5]$, but the drift times are similar for $M \geq 0.51$.

\begin{figure}[htbp]
    \centering
    \includegraphics[width=0.8\textwidth]{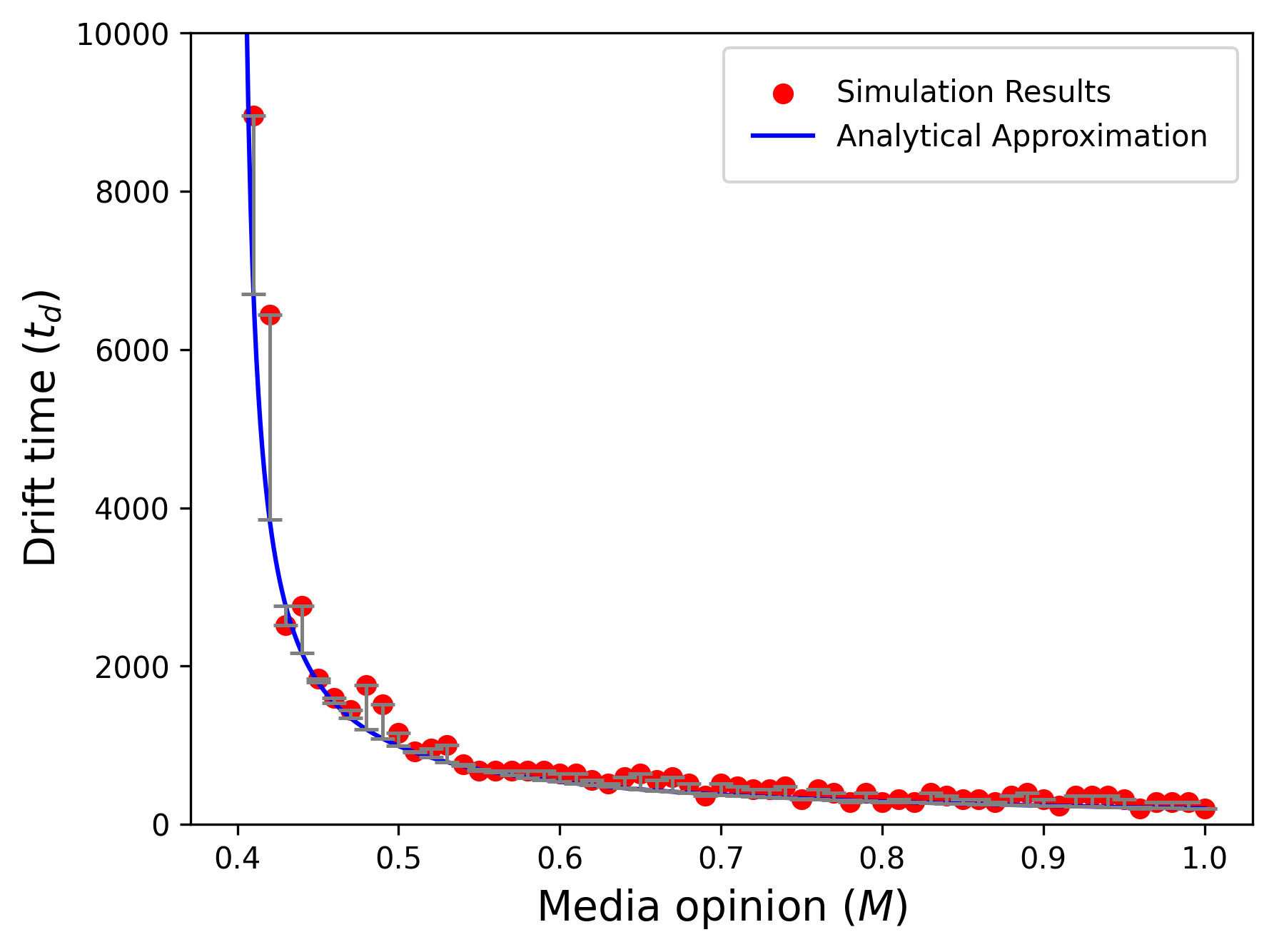}
    \caption{Comparison between our analytical approximation~\eqref{eq:drift_time} of the drift time and the results of numerical simulations of our ABM.
    We conduct {10} simulations for {each} media opinion between $M = 0.41$ and $M = 1$ in increments of $0.01$.}
    \label{fig:drift_time_comp}
\end{figure}


\subsubsection{Refinement of the Drift Equations \eqref{eq:drift_traj} and \eqref{eq:drift_time}}\label{subsubsec:refinement}

In Figures~\ref{fig:drift_trajectory_comp} and~\ref{fig:drift_time_comp}, we observed discrepancies between our simulations and our analytical approximations \eqref{eq:drift_traj} and \eqref{eq:drift_time}.
 We are particularly interested in the discrepancies between the asymptotes of the simulations and analytical approximations.
These discrepancies most likely result from our simplifying assumptions. {The opinion-cluster location $x_c$ is a function of time $t$ and media opinion $M$ and the drift time $t_d$ is a function of $M$, so we attempt} to reduce these discrepancies by adjusting $M$.

 By fitting the analytical asymptote $L$ (see Eq.~\eqref{eq:asymptote}) to match the numerically observed asymptote in Figure~\ref{fig:drift_trajectory_comp}, we obtain $M \approx 0.492$, which differs from the original value $M = 0.5$ by approximately $0.008$. Repeating this procedure for other values of $M$ reveals a consistent offset of roughly $0.008$. Given this offset, we introduce an ``effective media opinion'', which is the media opinion after accounting for systematic biases in our simplifying assumptions. The effective media opinion is $M' := M - \delta$, where $\delta = 0.008$. Substituting $M'$ (instead of $M$) into~\eqref{eq:drift_traj} and~\eqref{eq:drift_time} yields significantly improved approximations (see Figure~\ref{fig:drift_combined_improved}).

\begin{figure}[htbp]
    \centering
    
    \begin{subfigure}{0.8\textwidth}
        \centering
        \includegraphics[width=\textwidth]{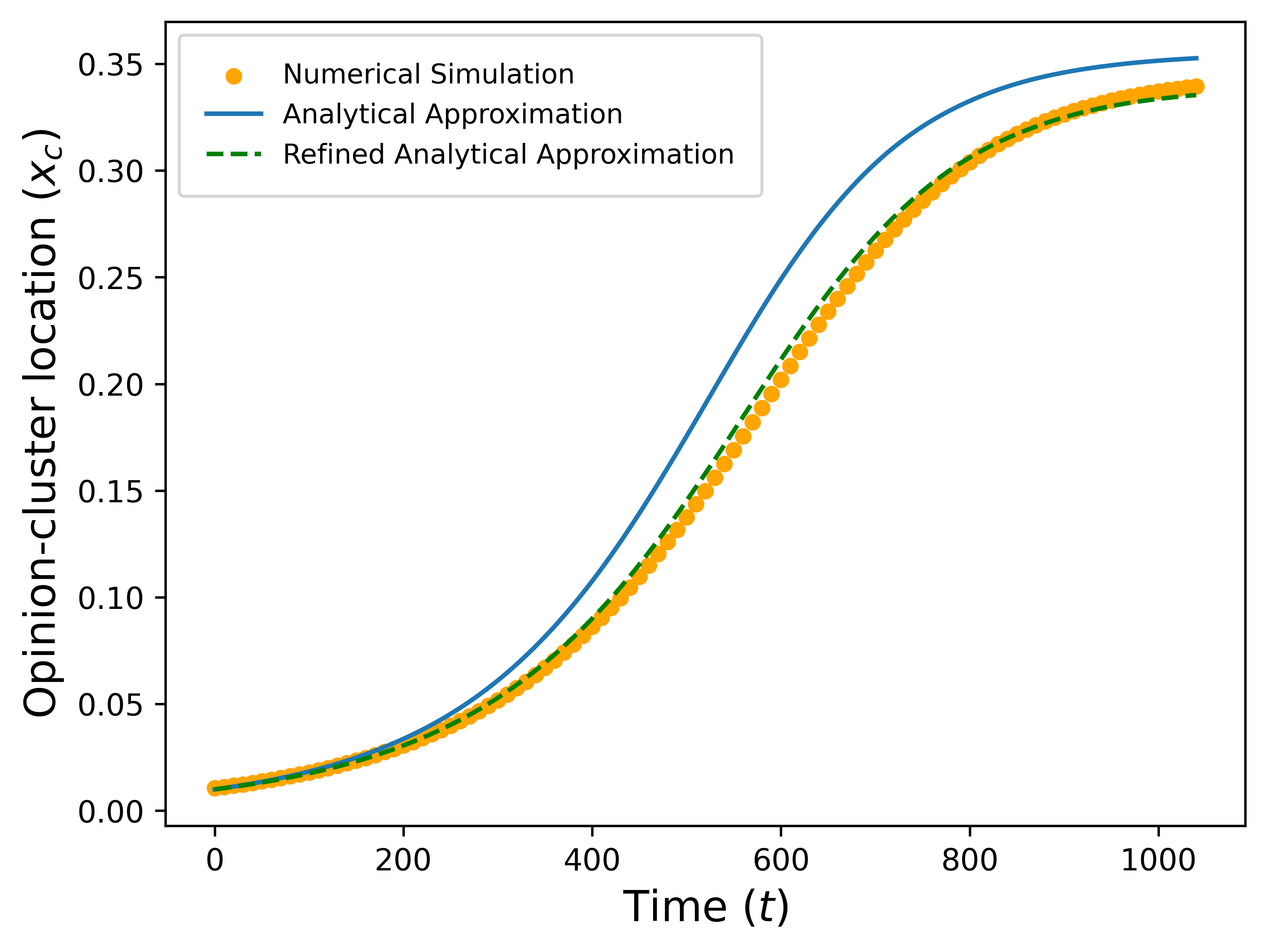}
        \caption{Refined analytical approximation of the drift trajectory.}
        \label{fig:drift_trajectory_improved}
    \end{subfigure}
    \hfill
    \begin{subfigure}{0.8\textwidth}
        \centering
        \includegraphics[width=\textwidth]{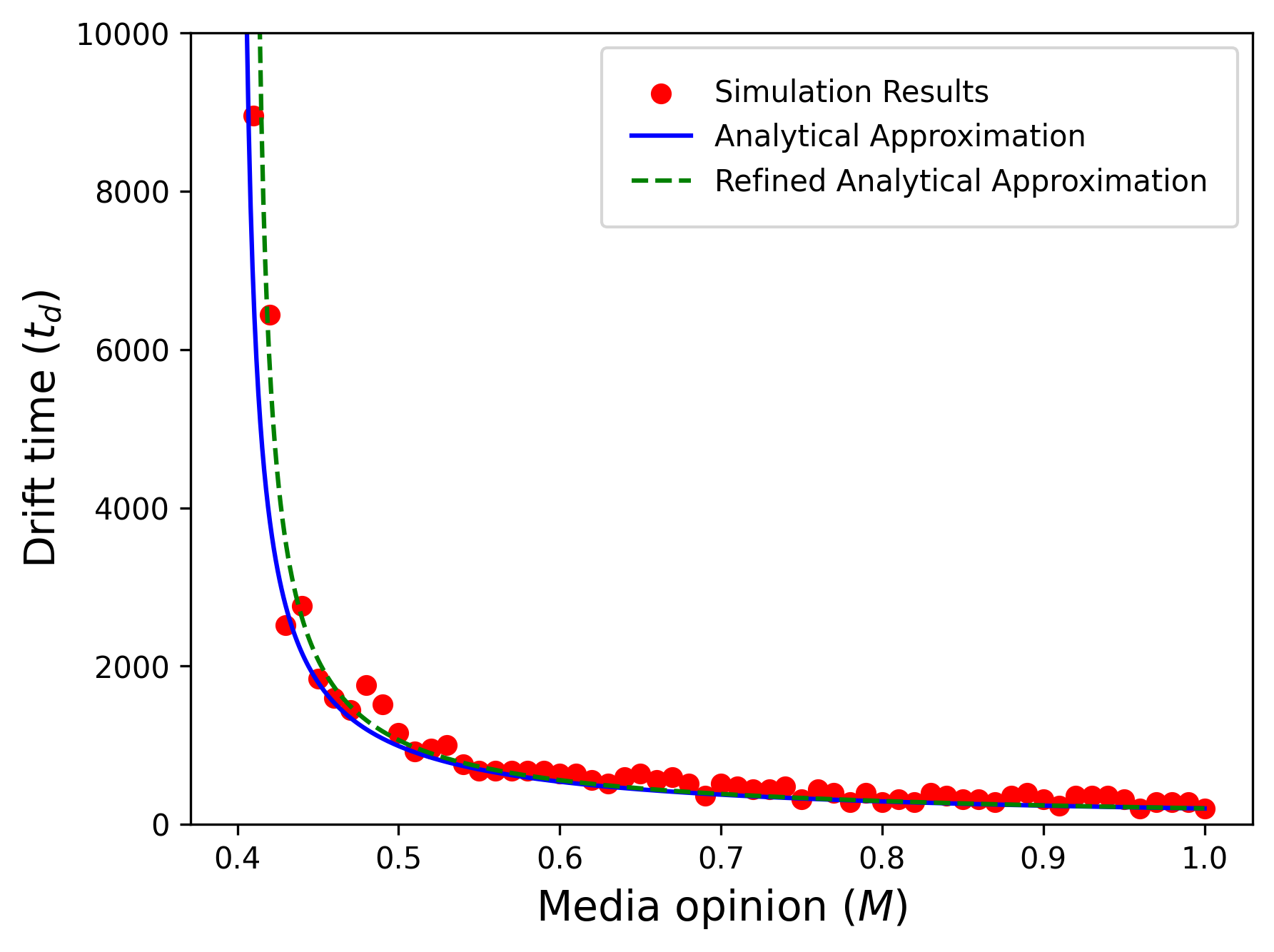}
        \caption{Refined analytical approximation of the drift time. The two curves are visually similar, but the original analytical approximation yields $R^2 \approx 0.8354$ and the refined approximation yields $R^2 \approx 0.9527$.}
        \label{fig:drift_time_improved}
    \end{subfigure}
    
    \caption{Refined analytical approximations of the drift behavior.
    }
    \label{fig:drift_combined_improved}
\end{figure}


\section{Conclusions and Discussion}\label{sec:conclusions}

{Studying agent-based models (ABMs) of opinion dynamics allows one to probe real-world behavior in idealized settings~\cite{starnini2025,volkening2025}.
In the present paper, we introduced an opinion model that examine the effects of media polarization and how it can drive opinion shifts in a population.}


\subsection{Summary of our results}\label{subsec:summary}

We augmented a bounded-confidence model (BCM) that is based on the Deffuant--Weisbuch (DW) framework by incorporating media sources. These media agents hold
opposing opinion values, $M$ and $-M$, in a one-dimensional opinion space. 

Using numerical simulations, we observed that the long-term behavior of the baseline DW model has opinion clusters that are located 
symmetrically about the center of the opinion space. However, in certain parameter regions, the media agents in our BCM break this symmetry and lead to
opinion drifts, with the dominant opinion cluster drifting from the center of opinion space toward one of the media agents. 

We then derived a continuous-time approximation of our ABM. In this {continuous-time} framework, we proved that opinion drift occurs only when the media opinion $M > 0.4$. 
The {continuous-time} description also allowed us to derive approximations for the stable opinion (i.e., the opinion value at which the dominant opinion cluster ceases drifting), the drift time (which is the time that is required for the dominant opinion cluster to move from the center of opinion space to the stable opinion), and the full trajectory of the dominant opinion cluster over time.

Comparing our {continuous-time} analytical results with the results of numerical simulations of our ABM revealed some discrepancies, particularly in the trajectory of the opinion drift. Because our approximation of the drift trajectory depends explicitly on $M$, we refined it by introducing a small constant offset $\delta$ to $M$. This refinement significantly improved our approximations.

Through our investigation, we demonstrated that a polarized media landscape can significantly affect the qualitative behavior of opinion dynamics in BCMs. 
Our findings suggest that the amount of media polarization plays a crucial role in these dynamics, as more media polarization leads to more polarized opinions in a population. 
More generally, our study contributes to the understanding of how external information sources and their associated opinions in a modern, highly connected society can help shape collective opinion dynamics.


\subsection{Outlook}\label{subsec:future}

There are many possible extensions of our work that can further understanding of media influence in BCMs and in opinion models more generally. 

One important avenue is to study the impact of network structure on the behavior of our ABM. In the present paper, we examined our ABM on complete graphs, but real social networks have diverse connection patterns, including community structure, core--periphery structure, and heterogeneous degree distributions. Different network structures may have different effects on both opinion drift 
and other qualitative dynamics of our ABM. Some relevant network structures to consider are Erd\H{o}s--R\'enyi graphs, stochastic block models, and random-graph models with heavy-tailed degree distributions~\cite{newman2018}.

There are also several interesting ways to modify our ABM. One relevant extension is to modify the media opinion values so that one media agent is more extreme than the other. Additionally, 
one can include distinct compromise parameters for interactions between two normal agents and interactions between normal agents and media agents. These extensions
introduce asymmetries, and studying their effects may provide insight into how such asymmetries influence opinion dynamics.

It is also important to study opinion drift in other opinion models. {In previous studies of {both BCMs and other ABMs}, Amblard and Deffuant~\cite{Amblard_Deffuant_2004}, Hegselmann and Krause~\cite{hegselmann2002opinion}, 
Sood et al.~\cite{sood2008drift}, and most recently {Ram\'{i}rez} et al.~\cite{Ramirez2024NonlinearVoter} observed opinion drifts that are similar to our opinion drifts.}
These observations suggest that opinion drift may arise from general mechanisms that {appear across} different interaction rules. It is important to clarify these mechanisms and to investigate
whether similar effects occur in opinion models with stubbornness, heterogeneous activity levels (e.g., see \cite{li2023}), and other features.
 One model to consider is the Friedkin--Johnson (FJ) model~\cite{friedkin1990social}, in which each agent has a
 compromise parameter that controls their stubbornness level. For example, one can consider a network with community structure and suppose, as in~\cite{WangXingJohansson2024FJRandomGraphs}, 
 that agents with different stubbornness levels are in different communities. One can then vary the compromise parameter and investigate whether opinion drift occurs and, if so, analyze its dependence on the stubbornness levels.
  Using a variety of different opinion models, it is also worth exploring {whether} mechanisms aside from the presence of stubborn agents can yield opinion drifts.


\section*{Acknowledgements} 

We thank Weiqi Chu for helpful discussions about writing code for our numerical simulations. We thank two anonymous people for helpful feedback on an earlier version of our manuscript. We used ChatGPT-4 to assist with the development and debugging of portions of our simulation code. We independently verified all output prior to integrating it into our workflow. Obviously, we assume responsibility for all content.





\end{document}